\documentclass[10pt,a4paper]{amsart}
\usepackage[all]{xy} %For commutative diagrams

\usepackage{amssymb} %For double arrows
\usepackage{eucal}

\usepackage{hyperref}

\numberwithin{equation}{section}
\newtheorem{definition}{Definition}[section]
\newtheorem{lemma}[definition]{Lemma}
\newtheorem{theorem}[definition]{Theorem}
\newtheorem{proposition}[definition]{Proposition}

\newtheorem{remarkth}[definition]{Remark}

\newenvironment{remark}{\begin{remarkth}\upshape}{\hfill$\diamond$\end{remarkth}}

\newcommand{\lcf}{\lbrack\! \lbrack}
\newcommand{\rcf}{\rbrack\! \rbrack}

\renewcommand{\d}{\mathrm{d}^\circ}

\newcommand{\Sec}[1]{\operatorname{Sec}(#1)}
\let\sec\Sec

\def\r{\ensuremath{\mathbb{R}}}% N\'{u}meros reales
\def\rk{{\mathbb R}^{k}}% Espacio euclideo de dimensi\'{o}n k
\def\tkq{T^1_kQ}
\def\tkqh{(T^1_k)^*Q}
\def\rktkq{\rk\times T^1_kQ}
\def\rktkqh{\rk\times (T^1_k)^*Q}
\def\rkq{\rk \times Q}
\def\rkq{\rk \times Q}

\def\ke{\stackrel{k}{\oplus} E}% suma directa de algebroides
\def\te{\mathcal{T}^E(\ke)}%prolongaci\'{o}n Lagrangiana
\def\keh{\stackrel{k}{\oplus} {E^{\,*}}}% suma directa de algebroides
\def\teh{\mathcal{T}^E(\keh)}%k-prolongaci\'{o}n Lagrangiana
\def\Lec{\mathop{Leg}} % Aplicaci\'{o}n de Legendre
\def\derpar#1#2{\displaystyle\frac{\partial{#1}}{\partial{#2}}}
\def\derpars#1#2#3{\displaystyle\frac{\partial^2{#1}}{\partial{#2}{\partial
{#3}}}}
\def\vf{\mathfrak{X}}% Campos de vectores
\def\d{{\rm d}}
\def\rke{\rk\times\ke}
\def\tec{\mathcal{T}^E(\rke)}
\def\tkec{(\mathcal{T}^E)^1_k(\rk\times\ke)}
\def\rkeh{\rk\times\keh}
\def\tech{\mathcal{T}^E(\rk\times\keh)}
\def\tkech{(\mathcal{T}^E)^1_k(\rk\times\keh)}

\def\qed{\ifvmode\Realemovelastskip\fi
{\unskip\nobreak\hfil\penalty50\hbox{}\nobreak\hfil \hbox{\vrule
height1.2ex width1.2ex}\parfillskip=0pt \finalhyphendemerits=0
\par\smallskip}}
\def\qed{\ifvmode\removelastskip\fi
{\unskip\nobreak\hfil\penalty50\hbox{}\nobreak\hfil \hbox{\vrule
height1.2ex width1.2ex}\parfillskip=0pt \finalhyphendemerits=0
\par\smallskip}}

\begin{document}

\title[$k$-cosymplectic
formalism on Lie algebroids]{Reduced classical field theories.\\ $k$-cosymplectic
formalism on Lie algebroids}

\author[D.\ Mart\'{\i}n de Diego]{D. Mart\'{\i}n de Diego}
\address{D.\ Mart\'{\i}n de Diego:
Instituto de Ciencias Matem\'aticas (CSIC-UAM-UCM-UC3M)\\ C/ Serrano
123, 28006 Madrid, Spain} \email{d.martin@imaff.cfmac.csic.es}

\author[S. Vilari\~no]{S. Vilari\~no}
\address{S. Vilari\~no:
Departamento de Matem\'{a}ticas, Facultad de Ciencias,
    Universidad de A Coru\~{n}a,
    Campus de A Zapateira
    15071-A Coru\~{n}a, Spain}
    \email{silvia.vilarino@udc.es}
%\address{Silvia Vilari\~no:
%Departamento de Matem\'{a}ticas
%    Universidade da Coru\~{n}a
%    15782-Santiago de Compostela, Spain}
%    \email{silvia.vilarino@usc.es}



\begin{abstract}
 In this paper we introduce a geometric description of Lagrangian and
Hamiltonian classical field theories on Lie algebroids in the framework
of $k$-cosymplectic geometry. We discuss the relation between Lagrangian
and Hamiltonian descriptions through a convenient notion of Legendre
transformation. The theory is a natural generalization of the standard
one; in addition, other interesting examples are studied, mainly on reduction of classical field theories.

\end{abstract}

%\thanks{This work has been partially supported by ....}

\maketitle

\begin{center}\today\end{center}

\tableofcontents

\section{Introduction}\label{intro}

The {\sl $k$-cosymplectic formalism}
is one of the simplest geometric framework for describing
many interesting cases of first-order classical field theories.
It is a generalization to field
theories of the standard cosymplectic formalism for non-autonomous mechanics and it is adequate for describing  field
theories with Lagrangians or Hamiltonians function explicitly depending    on coordinates in the basis or the set of parameters.  The foundation of the $k$-cosymplectic formalism
is the  $k$-cosymplectic manifolds \cite{LMORS-1998,{LMS-2001}}.

Historically, it is based on the so-called {\sl polysymplectic formalism}
developed by G\"{u}nther \cite{Gu-1987}, who introduced the
{\sl polysymplectic manifolds}. A refinement of this concept
allows us  to define {\sl $k$-symplectic manifolds} \cite{Aw-1992,{Aw-2000}}, which are
polysymplectic manifolds admitting {\sl Darboux-type coordinates} \cite{LMS-1988,{LMS-1993}}.
(Other different polysymplectic formalisms
for describing field theories have been also proposed
\cite{GMS-1997,{GMS-1999},Kana,McN-2000,Norris-1993,Sarda-1995}).

Sometimes, the Lagrangian and Hamiltonian functions are not defined on a $k$-cosymplectic manifold, for instance, in the reduction theory, where the reduced ``phase spaces" are not, in general, $k$-cosymplectic manifolds, even when the original phase space is a $k$-cosymplectic manifold. For instance, in this paper, we will see that when we consider reduction by symmetry of Lagrangian field theories, we obtain a reduced Lagrangian which can not described using the standard $k$-cosymplectic theory. In Mechanics this problem is solved using Lie algebroids instead of tangent and cotangent bundles (see \cite{LMM-2005, Weins-1996}).

The goal of this paper is to develop an extension of $k$-cosymplectic field theories  to Lie algebroids, such that, in the particular case $k=1$ we obtain the traditional  mechanics on Lie algebroids and when the Lie algebroid is the tangent bundle we derive the classical $k$-cosymplectic formalism. Classical field theories on Lie algebroids have already been studied in the literature. For instance, the multisymplectic formalism on Lie algebroids was presented in \cite{Mart-2004,{Mart-2005}}, the $k$-symplectic formalism on Lie algebroids was studied in \cite{LMSV-09}. In \cite{VC-06} a geometric framework for discrete field theories on Lie groupoids has been discussed.

The organization of the paper is as follows. In section \ref{EPeq} we summarize some aspects of the reduction on principal bundles developed by M.  Castrill\'{o}n {\it et al.} in \cite{CR-2003} and  \cite{CRS-2000}, the {\it covariant Lagrangian reduction}. This approach gives us examples of field theories on reduced  ``phase spaces'' which  are not, in general,  $k$-cosymplectic manifolds. Here we observe that is necessary to develop a theory more general than the $k$-cosymplectic formalism for field theory. In section \ref{gepre} we  recall some basic elements from the $k$-cosymplectic approach to first order classical field theories. In section \ref{algebroids} we remember some basic facts about Lie algebroids an the differential geometric aspects associated to them. In this section we also describe a particular example of Lie algebroid, called the {\it prolongation of a Lie algebroid over a fibration}. This Lie algebroid will be necessary for the further developments. In section \ref{cftla} the $k$-cosymplectic formalism is extended to the setting of Lie algebroids. The subsection \ref{CFTLA} describe the Lagrangian approach and the subsection \ref{Ha} describe the Hamiltonian approach. These formalisms are developed in an analogous way to the standard $k$-cosymplectic Lagrangian and Hamiltonian formalisms. We finish this section defining the Legendre transformation on the context of Lie algebroids and we establish the equivalence between both formalism, Lagrangian and Hamiltonian, when the Lagrangian function is hyperregular. In section \ref{examples} we show some examples where the theory can be applied.

 All manifolds and maps are $C^\infty$. Sum over crossed repeated
indices is understood. Along this paper one $k$-tuple of elements will be denoted by a bold symbol.

\section{Motivating example: Principal bundle reduction, covariant Euler-Poincar\'{e} equations.}\label{EPeq}

Reduction by symmetry of Lagrangian field theories is useful for  the implementation of many diverse mathematical models from geometric mechanics. One of the main approaches has been develop by M. Castrill\'{o}n-L\'{o}pez {\it et al.} in \cite{CR-2003}  and  \cite{CRS-2000} and it is referred  as {\it covariant Lagrangian reduction}.

The  papers on covariant Lagrangian reduction, \cite{CRS-2000}, dealt with the extension of classical Euler-Poincar\'{e} reduction of variational principles to the field theoretic context, the idea of this paper is the following: a field theory was formulated on a principal bundle and was reduced by the structure group. This process can be summarized as follows. We begin with a right principal bundle $\pi\colon P\to M$ with structure group $G$. The group $G$ naturally acts  on $J^1P$ by $(j^1_xs)\cdot g\mapsto j^1_x(R_g\circ s)$, for any $j^1_xs\in J^1P$ and $g\in G$. One considers a Lagrangian $L\colon J^1P\to \r$, invariant under the natural action induced by the structure group $G$. The reduced variational problem now takes place on $\mathcal{C}(P) =(J^1P)/G$, the bundle of connections, for more details see \cite{CRS-2000}.

Now we consider the following particular case:  $P=\rk\times G$ and $M=\rk$, that is, the trivial bundle $\rk\times G\to \rk$, in this case, $J^1P$ can be identified with $\rk\times T^1_kG$, where, $T^1_kG$ is the tangent bundle of $k^1$-velocities of $G$, that is, the Whitney sum of $k$ copies of $TG$ (see section \ref{gepre}).

Let $L\colon J^1P\equiv\rk\times T^1_kG\to \r$ be a Lagrangian invariant under the natural action of $G$ on $\rk\times T^1_kG$. In this case, we make the identifications
$$J^1(\rk\times G)/G\cong(\rk\times T^1_kG)/G\cong(\rk\times G\times \mathfrak{g}\times\stackrel{k}{\ldots}\times \mathfrak{g})/G\cong\rk\times \mathfrak{g}\times\stackrel{k}{\ldots}\times \mathfrak{g}$$ and then
 the reduced Lagrangian is a function $l\colon (J^1P)/G\equiv (\rk\times T^1_kG)/G\equiv \rk\times (\mathfrak{g}\oplus \stackrel{k}{\ldots}\oplus \mathfrak{g})\to \r$.

Let us observe that in this simple example of covariant Lagrangian reduction, the reduced Lagrangian is not defined on a $k$-cosymplectic manifold.

In this paper we will study classical field theories on Lie algebroids using the $k$-cosymplectic approach. In this setting the above example can be solved. Furthermore, we will develop  a framework that:
\begin{enumerate}
\item Reduces the classical $k$-cosymplectic field theories, \cite{LMORS-1998,LMS-2001} to particular cases.
\item Reduces mechanics on Lie algebroids, see for instance \cite{CLMMM-2006,Mart-2001}, to  particular cases.
\end{enumerate}

\section{Geometric preliminaries}\label{gepre}

In this section we recall some basic elements from the $k$-cosymplectic approach to classical field theories \cite{LMORS-1998,{LMS-2001}}.

\subsection{The  manifold $\r^k\times (T^1_k)^*Q$}

Let $Q$ be an $n$-dimensional differentiable manifold  and
${\pi}_{Q}: T^{*}Q \to Q$ its cotangent bundle.
We denote by $(T^1_k)^{\;*}Q$ the Whitney sum $T^{*}Q \oplus \stackrel{k}{\dots}
\oplus T^{*}Q$ of $k$ copies of $T^{*}Q$

 $(T^1_k)^{\;*}Q$ can be  identified with the manifold
$J^1(Q,\rk)_\mathbf{0}$ of $k^1$-covelocities of  $Q$, that is, 1-jets of maps $\sigma\colon Q\to \rk$ with target
at $\mathbf{0}\in \rk$,say
\[
\begin{array}{ccc}
J^1(Q,\r^k)_\mathbf{0} & \equiv & T^{*}Q \oplus \stackrel{k}{\dots} \oplus T^{*}Q \\
j^1_{q,\mathbf{0}}\sigma  & \equiv & (d\sigma^1(q), \dots ,d\sigma^k(q))\ ,
\end{array}
\]
where $\sigma^A= pr_A \circ \sigma:Q \longrightarrow \r$ is the
$A{th}$ component of $\sigma$ and  $pr_A:\r^k \to \r$ are the
canonical projections, $1\leq A\leq k$. For this reason, $(T^1_k)^{\;*}Q$ is also called {\sl the bundle of
$k^1$-covelocities of the manifold $Q$}.

The manifold $J^1\hat{\pi}_{Q}$ of  1-jets of sections of the trivial
bundle $\widehat{\pi}_{Q}:\rk \times Q \to Q$ is diffeomorphic to $\rk \times
(T^1_k)^{\;*}Q$, via the diffeomorphism given by
\[
\begin{array}{rcl}
J^1\widehat{\pi}_{Q} & \longrightarrow & \rk   \times (T^1_k)^{\;*}Q \\
\noalign{\medskip} j^1_q\phi= j^1_q(\phi_{\rk},Id_{Q})  & \longmapsto &
(\phi_{\rk}(q), \alpha^1_q, \ldots ,\alpha^k_q) \ ,
\end{array}
\]
where $\phi_{\rk}: Q \stackrel{\phi}{\to}  \rkq
\stackrel{\widehat{\pi}_{\rk}}{\to}\rk $,
$1\leq A \leq k$, $(\phi_{\rk})^A:Q\stackrel{\phi_{\rk}}{\to} \rk
\stackrel{pr_A}{\to} \r$ and $\alpha^A_q=d(\phi_{\rk})^A(q)$.

Throughout all the paper we use the following notation for the canonical projections
\[\xymatrix@C=13mm{\rk\times (T^1_k)^*Q\ar[r]^-{({\widehat{\pi}}_Q)_{1,\,0}}\ar[dr]_-{({\widehat{\pi}}_Q)_1}
& \rkq\ar[d]^-{{\widehat{\pi}}_Q}\\
 & Q
}\]
  where
$$\widehat{\pi}_Q(\mathbf{t},q)=q, \quad (\widehat{\pi}_Q)_{1,0}(\mathbf{t},\alpha^1_q, \ldots
,\alpha^k_q)=(\mathbf{t},q), \quad (\widehat{\pi}_Q)_1(\mathbf{t},\alpha^1_q, \ldots
,\alpha^k_q)=q \, ,$$ with $\mathbf{t}\in \rk $, $q\in Q$ and $(\alpha^1_q, \ldots ,\alpha^k_q)\in (T^1_k)^{\;*}Q$.

If $(q^i)$ are local coordinates on $U \subseteq Q$, then the
induced local coordinates $(q^i , p_i)$, $1\leq i \leq n$, on
$({\pi}_Q)^{\,-1}(U)=T^{*}U\subset T^{*}Q$, are expressed by
$$
q^i(\alpha_q)=q^i(q), \quad p_i(\alpha_q)= \alpha_q \left(
\frac{\partial}{\partial q^i}\Big\vert_q \right)\, ,
$$
and the induced local coordinates  $(t^A,q^i ,
p^A_i),\; 1\leq i\leq n ,\, 1\leq A\leq k,$ on $[(\widehat{\pi}_Q)_1]^{\,-1}(U)=\rk \times (T^1_k)^{\;*}U$ are given
by
$$
t^A(\mathbf{t},\alpha^1_q, \ldots ,\alpha^k_q) = t^A\, , \quad
q^i(\mathbf{t},\alpha^1_q, \ldots ,\alpha^k_q) = q^i(q)\, , \quad
p_A^i(\mathbf{t},\alpha^1_q, \ldots ,\alpha^k_q) =
\alpha^A_q\left(\displaystyle\frac{\partial}{\partial q^i}\Big\vert_q
\right)\, .
$$

On $\rk\times (T^1_k)^{\;*}Q$, we consider the differential forms
$$
 \eta^A\equiv dt^A=(\pi^A_1)^*dt^A\, , \quad \theta^A=
(\pi^A_2)^{\;*}\theta\, , \quad \omega^A=
(\pi^A_2)^{\;*}\omega\, ,
$$
 $\pi^A_1:\rk \times (T^1_k)^{\;*}Q \rightarrow \r$ ,
$\pi^A_2:\rk \times (T^1_k)^{\;*}Q \rightarrow T^{\;*}Q$ being the canonical
projections defined by
$$
\pi^A_1(\mathbf{t},\alpha^1_q, \ldots ,\alpha^k_q)=t^A \,,\quad
\pi^A_2(\mathbf{t},\alpha^1_q, \ldots ,\alpha^k_q)=\alpha^A_q\, ,
$$
where $\omega=-d\theta=dq^i \wedge dp_i$ is the canonical symplectic form
on $T^{*}Q$ and $\theta=p_i \, dq^i$ is the Liouville $1$-form on
$T^{*}Q$. Obviously $\omega^A = -d\theta^A,\; 1\leq A\leq k$.

In local coordinates we have
\begin{equation}\label{locexp}
    \theta^A = \displaystyle \, p^A_i
dq^i \quad  ,  \quad \omega^A = \displaystyle  dq^i \wedge dp^A_i\, .
\end{equation}

Moreover, let       $$ V^{\;*}=ker\, \left( \, ((\hat{\pi}_{Q})_{1,0}
)_*\right)=\left\langle\frac{\displaystyle\partial}
{\displaystyle\partial p^1_i}, \dots,
\frac{\displaystyle\partial}{\displaystyle\partial
p^k_i}\right\rangle_{i=1,\ldots , n}\; ,
$$
be the vertical distribution of the bundle
$(\hat{\pi}_{Q})_{1,0}:\rk\times (T^1_k)^{\;*}Q \to \rkq$.

A simple inspection of the expressions in local coordinates
(\ref{locexp}) shows that the forms $\eta^A$ and $\omega^A$ are
closed, and the following relations hold
\begin{enumerate}
\item $\eta^1\wedge\dots\wedge \eta^k\neq 0$,\quad
$(\eta^A)\vert_{V^{\;*}}=0,\quad (\omega^A)\vert_{V^{\;*}\times
V^{\;*}}=0,$

\item $(\displaystyle{ {\cap_{A=1}^{k}} \ker \eta^A}) \cap (\displaystyle
{\cap_{A=1}^{k}} \ker \omega^A)=\{0\}$, \quad $dim(\displaystyle
{\cap_{A=1}^{k}} \ker \omega^A)=k,$
\end{enumerate}

{}From the above  geometrical model, the following definition is
introduced  in \cite{LMORS-1998}:

\begin{definition}\label{deest}
Let $M$ be a differentiable manifold  of dimension $k(n+1)+n$. A
{\rm $k$--cosymplectic structure} on $M$ is a family
$(\eta^A,\omega^A,V;1\leq A\leq k)$, where each $\eta^A$ is a closed
$1$-form, each $\omega^A$ is a closed $2$-form and $V$ is an
integrable  $nk$-dimensional distribution on $M$, satisfying $(i)$ and
$(ii)$.

$M$ is said to be an {\rm $k$--cosymplectic manifold}.
\end{definition}

The following theorem has been proved in  \cite{LMORS-1998}.

\begin{theorem}  {\rm (Darboux Theorem):}
If $M$ is   an  $k$--cosymplectic manifold, then around each
  point of $M$ there exist local coordinates
$(t^A,q^i,p^A_i)$ such that
$$
\eta^A=dt^A,\quad \omega^A=dq^i\wedge dp^A_i, \quad
V=\left\langle\frac{\displaystyle\partial} {\displaystyle\partial
p^1_i}, \dots, \frac{\displaystyle\partial}{\displaystyle\partial
p^k_i}\right\rangle_{i=1,\ldots , n}.
$$
\end{theorem}

In consequence, the canonical model for these geometrical structures is   $(\rk
\times (T^1_k)^{\;*}Q,\eta^A,\omega^A,V^{\;*})$. See, for instance \cite{LMORS-1998,LMS-2001,MSV-2005}

\subsection{The manifold $\rk \times T^1_kQ$}

 Let $Q$ be an $n$-dimensional manifold and $\tau_Q\colon TQ\to Q$ its tangent bundle. We denote by  $\tkq$ the Whitney sum $TQ
\oplus\stackrel{k}{\ldots}\oplus TQ$ of $k$ copies of $TQ$, with projection $\tau^k_Q\colon\tkq\to Q,\,\tau^k_Q({v_1}_q,\ldots, {v_k}_q)=q$, where ${v_A}_q\in T_qQ,\, A=1,\ldots, k$.
$\tkq$ can be identified with the manifold $J_\mathbf{0}^1(\rk,Q)$ of  $k^1$-velocities of $Q$, that
is  1-jets of maps $\sigma\colon\rk\to Q$   with the source at $\mathbf{0}\in\rk$, say
$$\begin{array}{ccc}
J_\mathbf{0}^1(\rk,Q) & \equiv & TQ\,
\oplus\stackrel{k}{\ldots}\oplus\, TQ\\\noalign{\medskip}
j^1_{\mathbf{0},q}\sigma &\equiv & ({v_1}_q,\ldots, {v_k}_q)
\end{array}$$
 where $q=\sigma(\mathbf{0})$  and ${v_A}_q=T_{\mathbf{0}}\sigma\left(\frac{\partial}{\partial t^A}\Big|_{\mathbf{t}=\mathbf{0}}\right), (t^1,\ldots, t^k)$ being the standard coordinates on $\rk$.  $\tkq$ is called {\it the tangent
bundle of $k^1$-velocities of Q} (see \cite{Morimoto-1970}).

The manifold $J^1\hat{\pi}_{\rk}$ of 1-jets of sections of the trivial
bundle $\hat{\pi}_{\rk}:\rk \times Q \to \rk$ is diffeomorphic to $\rk
\times  T^1_kQ$, via the diffeomorphism given by
$$
\begin{array}{rcl}
J^1\hat{\pi}_{\rk} & \to & \rk   \times T^1_kQ \\
\noalign{\medskip} j^1_t\phi= j^1_t(Id_{\rk},\phi_Q) & \to & (
t,v_1, \ldots ,v_k)
\end{array}
$$
where $\phi_Q: \rk \stackrel{\phi}{\to}  \rkq \stackrel{\hat{\pi}_{Q}}{\to}Q
$, and
$$
v_A=T_{\mathbf{t}}\phi_Q\left(\displaystyle\frac{\partial}{\partial
t^A}\Big\vert_\mathbf{t}\right)\, , \quad  1\leq A \leq k \, .
$$

Let $p_Q:\rk\times \tkq \to Q$ be the  canonical projection. If $(q^i)$ are local coordinates on $U
\subseteq Q$,  then the induced local coordinates  $(t^A,q^i , v^i_A)$ on
$p_Q^{-1}(U)=\rk \times T^1_kU$ are expressed by
$$t^A(\mathbf{t},{v_1}_q,\ldots , {v_k}_q)   =   t^A; \quad
q^i(\mathbf{t},{v_1}_q,\ldots , {v_k}_q) =q^i(q); \quad
v_A^i(\mathbf{t},{v_1}_q,\ldots , {v_k}_q)   =
\langle dq^i, v_A\rangle\, , $$ where $1\leq i \leq n,\, 1\leq A \leq
k$.

Throughout the paper we use the following notation for the canonical
projections
\[\xymatrix@C=13mm{\rk\times T^1_kQ\ar[r]^-{(\hat{\pi}_{\r^k})_{1,\,0}}\ar[dr]_-{(\hat{\pi}_{\r^k})_1}
& \rkq\ar[d]^-{\hat{\pi}_{\r^k}}\\
 &\r^k
}\]
 where, for $\mathbf{t}\in \rk $, $q\in Q$ and $({v_1}_q,\ldots , {v_k}_q)\in
T^1_kQ$,
$$\hat{\pi}_{\rk}(\mathbf{t},q)=\mathbf{t}, \quad (\hat{\pi}_{\rk})_{1,0}(\mathbf{t},{v_1}_q,\ldots ,
{v_k}_q)=(\mathbf{t},q), \quad (\hat{\pi}_{\rk})_1(\mathbf{t},{v_1}_q,\ldots , {v_k}_q)=\mathbf{t}  \ .$$

\subsection{k-vector fields and integral sections}\ \label{212}

Let $M$ be an arbitrary manifold.

\begin{definition}  \label{kvector} A section ${\bf X} : M \longrightarrow T^1_kM$ of the projection
$\tau^k_M$ will be called a  {\rm $k$-vector field} on $M$.
\end{definition}

To give a $k$-vector field ${\bf X}$ is equivalent to give a
family of $k$ vector fields $X_{1}, \dots, X_{k}$. Hence in the sequel we will indistinctly write $\mathbf{X}=(X_1, \ldots, X_k)$.

\begin{definition} \label{integsect}
An {\rm integral section} of the $k$-vector field  \, $\mathbf{X}=(X_{1},
\dots, X_{k})$, passing through a point $x\in M$,  is a map
$\psi:U_\mathbf{0}\subset \r^k \rightarrow M$, defined on some neighborhood
$U_\mathbf{0}$ of $\mathbf{0}\in \rk$, such that $\psi(\mathbf{0})=x$, and
\begin{equation}\label{gintdec}
\psi_{*}(\mathbf{t})\left(\displaystyle\frac{\displaystyle\partial}{\displaystyle\partial
t^A}\Big\vert_\mathbf{t}\right) = X_{A}(\psi (\mathbf{t})) \ , \quad \mbox{for every}
\quad \mathbf{t}\in U_\mathbf{0},\, 1\leq A\leq k,
\end{equation}
or,  equivalently,  $
\psi({\mathbf{0}})= {x}$ and $\psi$ satisfies
${  \mathbf{X}}\circ\psi=\psi^{(1)}$, where  $\psi^{(1)}$ is the first
prolongation of $\psi$  to $T^1_kM$, defined by
$$
\begin{array}{rccl}\label{1prolong}
\psi^{(1)}: & U_{0}\subset \r^k & \longrightarrow & T^1_kM
\\\noalign{\medskip}
 &\mathbf{t} & \longrightarrow & \psi^{(1)}(\mathbf{t})=j^1_{\mathbf{0}}\psi_\mathbf{t}\equiv
 \left(\psi_*(\mathbf{t})\left(\derpar{}{t^1}\Big\vert_\mathbf{t}\right),\ldots,
\psi_*(\mathbf{t})\left(\derpar{}{t^k}\Big\vert_\mathbf{t}\right)\right) \, ,
 \end{array}
$$ where $\psi_\mathbf{t}({  \mathbf{s}})=\psi(\mathbf{t}+{  \mathbf{s}})$.

A $k$-vector field $\mathbf{X}=(X_1,\ldots , X_k)$ on $M$ is said to be {\rm
integrable} if there is an integral section passes through every
point of $M$.
\end{definition}

\begin{remark}{\rm In the $k$-cosymplectic formalism,  the solutions of Euler-Lagrange's field equations are the integral
 sections of $k$-vector fields. In the case $k=1$, this definition coincides with
the classical definition of the integral curve of a vector field. }\end{remark}

\section{Lie algebroids}\label{algebroids}

In this section we present some basic facts about Lie algebroids,
including features of the associated differential calculus and results on Lie algebroid morphisms that will be
necessary. For further information on groupoids and Lie algebroids, and their roles in differential geometry, see \cite{CW-1999,HM-1990,Mack-1987,Mack-1995}.

\subsection{Lie algebroid: definition}

Let $E$ be a vector bundle of rank $m$ over a manifold $Q$ of
dimension $n$, and let $\tau: E\to Q$ be the vector bundle projection.
Denote by ${\rm Sec}(E)$ the $C^\infty(Q)$-module of sections of
$\tau$. A {\it Lie algebroid structure}
$(\lcf\cdot,\cdot\rcf_E,\rho_E)$ on $E$ is a Lie bracket
$\lcf\cdot,\cdot\rcf_E: {\rm Sec}(E)\times {\rm Sec}(E)\to {\rm Sec}(E)$ on
the space ${\rm Sec}(E)$, together with  an {\it anchor map} $\rho_E: E\to
TQ$ and its, identically denoted, induced $C^\infty(Q)$-module homomorphism
$\rho_E:{\rm Sec}(E)\to \vf{(Q)}$ such that the {\it compatibility condition}
\[
\lcf\sigma_1,f\sigma_2\rcf_E=
f\lcf\sigma_1,\sigma_2\rcf_E+(\rho_E(\sigma_1)f)\sigma_2\,,\]
holds for any smooth functions $f$ on $Q$ and sections $\,\sigma_1,\sigma_2$ of $E$ (here $\rho_E(\sigma_1)$ is the vector
field on $Q$ given by
$\rho_E(\sigma_1)( {q})=\rho_E(\sigma_1( {q}))$). The triple
$(E,\lcf\cdot,\cdot\rcf_E,\rho_E)$ is called a {\it Lie algebroid over $Q$}.
{}From the compatibility condition and the Jacobi identity, it follows
that  $\rho_E:{\rm Sec}(E)\to\vf{(Q)}$ is a homomorphism
between the Lie algebras $({\rm Sec}(E),\lcf\cdot,\cdot\rcf_E)$ and
$(\vf{(Q)},[\cdot,\cdot])$. The following are examples of Lie algebroids.
\begin{enumerate}
\item {\bf Real Lie algebras of finite dimension}. Any real Lie algebra of finite dimension is a Lie algebroid over a single point.

\item {\bf The tangent bundle.} If $TQ$ is the tangent bundle of a
manifold $Q$, then, the triple $(TQ,[\cdot,\cdot],id_{TQ})$ is a Lie
algebroid over $Q$, where $id_{TQ}:TQ\to TQ$ is the identity map.

\item Another important  example of a Lie algebroid may be constructed as follows.
Let $\pi: P\to Q$ be a principal bundle with structural group $G$.
Denote by $\Phi:G\times P\to P$ the free action of $G$ on $P$ and
by $T\Phi:G\times TP\to TP$ the tangent action of $G$ on $TP$.
Then the sections of the quotient vector bundle
$\tau_{P|G}: TP/G\to Q=P/G$
may be identified with the vector fields on $P$ which are
invariant under the action $\Phi$. Since every $G$-invariant
vector field on $P$ is $\pi$-projectable and the
standard Lie bracket on vector fields is closed with respect to
$G$-invariant vector fields, we can define a Lie algebroid
structure on $TP/G$. This Lie algebroid over $Q$ is called
{\bf the Atiyah (gauge) algebroid associated with the principal
bundle} $\pi:P\to Q$   \cite{LMM-2005,Mack-1987}.

\end{enumerate}

Throughout this paper, the role played by a Lie algebroid is the same as the  tangent bundle of
$Q$. In this way, one regards an element $e$ of $E$ as a generalized
velocity, and the actual velocity ${  v}$ is obtained when we apply the
anchor map to ${  e}$, i.e. ${  v}=\rho_E({  e})$.

Let $(q^i)_{i=1}^n$ be local coordinates on a neighborhood $U$ of $Q$ and
$\{e_\alpha\}_{1\leq \alpha\leq m}$  a local basis of sections of $\tau$.
Given ${e}\in E$ such that $\tau({e})= {q}$, we can write
$e=y^\alpha(e)e_\alpha( {q})\in E_ {q}$, i.e. each
section $\sigma$ is  given locally by $\sigma\big\vert_{U}=y^\alpha
e_\alpha$ and the
coordinates of ${  e}$ are $(q^i({e}),y^\alpha({e}))$. A Lie algebroid structure on $Q$ is locally determined  as a set of
local {\it structure
functions}  $\rho^i_\alpha,\; \mathcal{C}^\gamma_{\alpha\beta}$
on $Q$ that are defined by
\begin{equation}\label{structure} \rho_E(e_\alpha)=\rho^i_\alpha
\displaystyle\frac{\partial}{\partial q^i} ,\quad
\lcf e_\alpha,e_\beta\rcf_E=\mathcal{C}^\gamma_{\alpha\, \beta}e_\gamma\,
\end{equation} and
satisfy the  relations
\begin{equation}\label{ecest}\displaystyle\sum_{cyclic(\alpha,\beta,\gamma)}\left(\rho^i_\alpha\displaystyle\frac{\partial
\mathcal{C}^\nu_{\beta\gamma}}{\partial
q^i}+\mathcal{C}^\nu_{\alpha\mu}\mathcal{C}^\mu_{\beta\gamma}\right)=0\;
\quad , \quad  \rho^j_\alpha\displaystyle\frac{\partial \rho^i_\beta}{\partial
q^j}- \rho^j_\beta\displaystyle\frac{\partial \rho^i_\alpha}{\partial
q^j}=\rho^i_\gamma \mathcal{C}^\gamma_{\alpha\beta}\;.
\end{equation}These relations, which are a consequence of the compatibility
condition and Jacobi's identity,  are usually called {\it the structure equations} of the Lie algebroid $E$.

\subsection{Exterior differential}
A Lie algebroid structure on
$E$ allows us to define {\it the exterior differential of $E$},
$\d^E:{\rm Sec}(\bigwedge^l E^*)\to {\rm Sec}(\bigwedge^{l+1} E^*)$,
as follows:
{\small\begin{equation}\label{difE}\begin{array}{lcl}
\d^E\mu(\sigma_1,\ldots, \sigma_{l+1})&=&
\displaystyle\sum_{i=1}^{l+1}(-1)^{i+1}\rho_E(\sigma_i)\mu(\sigma_1,\ldots,
\widehat{\sigma_i},\ldots, \sigma_{l+1})\\\noalign{\medskip} &+&
\displaystyle\sum_{i<j}(-1)^{i+j}\mu([\sigma_i,\sigma_j]_E,\sigma_1,\ldots,
\widehat{\sigma_i},\ldots, \widehat{\sigma_j},\ldots
\sigma_{l+1})\;,
\end{array}\end{equation}}\noindent for $\mu\in {\rm Sec}(\bigwedge^lE^*)$ and
$\sigma_1,\ldots,\sigma_{l+1}\in {\rm Sec}(E)$. It follows that $\d^E$
is a cohomology operator, that is, $(\d^E)^2=0$.

In particular, if $f:Q\to\r$ is a  smooth real function then
$\d^E f(\sigma)=\rho_E(\sigma)f$, for $\sigma\in {\rm Sec}(E)$. Locally, the
exterior differential is determined by
\begin{equation}\label{difEloc}\d^Eq^i=\rho^i_\alpha e^\alpha\quad \makebox{and}\quad
\d^Ee^\gamma=-\displaystyle\frac{1}{2}\mathcal{C}^\gamma_{\alpha\beta}e^\alpha\wedge
e^\beta\,,\end{equation}where $\{e^\alpha\}$ is the dual basis of
$\{e_\alpha\}$.

The usual Cartan calculus extends to the case of Lie algebroids: for
every section $\sigma$ of $E$ we  have a derivation $\imath_\sigma$
(contraction) of degree $-1$ and a derivation
$\mathcal{L}_\sigma=\imath_\sigma\circ d + d\circ \imath_\sigma$
(the Lie derivative) of degree $0$; for more details, see
\cite{Mack-1987,Mack-1995}.

\subsection{Morphisms}
 Let $(E,\lcf\cdot,\cdot\rcf_E,\rho_E)$ and
$(E',\lcf\cdot,\cdot\rcf_{E'},\rho_{E'})$ be two Lie algebroids over $Q$ and
$Q'$ respectively, and suppose that
$\Phi=(\overline{\Phi},\underline{\Phi})$ is a vector bundle map,
that is $\overline{\Phi}:E\to E'$ is a fiberwise linear map over
$\underline{\Phi}:Q\to Q'$. The pair
$(\overline{\Phi},\underline{\Phi})$ is said to be a {\it Lie
algebroid morphism} if \begin{equation}\label{lie morph}\d^E
(\Phi^*\sigma')=\Phi^*(\d^{E'}\sigma ')\,,\quad\makebox{ for all }
\sigma '\in Sec(\textstyle\bigwedge^l (E')^*)\makebox{ and for all
$l$.}\end{equation}

Here $\Phi^*\sigma '$ is the section of the vector
bundle $\bigwedge^l E^*\to Q$ defined (for $l>0$) by
\begin{equation}\label{pullsec}(\Phi^*\sigma ')_ {q}({  e}_1,\ldots,{  e}_l) = \sigma_{\underline{\Phi}
( {q})}' (\overline{\Phi}({  e}_1),\ldots,
\overline{\Phi}({  e}_l))\,,\end{equation} for $ {q}\in Q$ and ${  e}_1,\ldots,
{  e}_l\in E_ {q}$.  In  particular,   when $Q=Q'$ and
$\underline{\Phi}=id_Q$ then (\ref{lie morph}) holds if and only if
\[\lcf\overline{\Phi}\circ\sigma_1,\overline{\Phi}\circ\sigma_2\rcf_{E'} =
\overline{\Phi}\lcf\sigma_1,\sigma_2\rcf_E,\quad
\rho_{E'}(\overline{\Phi}\circ\sigma)=\rho_E(\sigma),\quad \makebox{for }
\sigma,\sigma_1,\sigma_2\in Sec(E)\,.\]

Let $(q^i)$ be a local coordinate system on $Q$ and $(\bar{q}^i)$ a
local coordinate system on $Q'$. Let $\{e_\alpha\}$ and
$\{\bar{e}_{\bar{\alpha}}\}$ be local base of sections of $E$ and $E'$,
respectively, and $\{e^\alpha\}$ and $\{\bar{e}^{\bar{\alpha}}\}$ their respective dual
base. The vector bundle map $\Phi$ is determined by the relations
$\Phi^*{\bar{q}^{\bar{i}}}=\phi^{\bar{i}}( {q})$ and $\Phi^* \bar{e}^{\bar{\alpha}}=
\phi^{\bar{\alpha}}_\beta e^\beta$ for certain local functions $\phi^{\bar{i}}$ and
$\phi^{\bar{\alpha}}_\beta$ on $Q$. In this coordinate system
$\Phi=(\overline{\Phi},\underline{\Phi})$ is a   Lie
algebroid morphism if and only if
\begin{equation}\label{morp cond}
  (\rho_E)^j_\alpha \displaystyle\frac{\partial \phi^{\bar{i}}}{\partial q^j} =
  (\rho_{E'})^{\bar{i}}_{\bar{\beta}}\phi^{\bar{\beta}}_\alpha\quad , \quad
\phi^{\bar{\beta}}_\gamma\mathcal{C}^\gamma_{\alpha\delta}
 =\left((\rho_E)^i_\alpha\displaystyle\frac{\partial \phi^{\bar{\beta}}_\delta}{\partial
q^i} - (\rho_E)^i_\delta \displaystyle\frac{\partial \phi^{\bar\beta}_\alpha}{\partial
q^i}\right) +
{\bar{\mathcal{C}}}^{\bar{\beta}}_{\bar{\theta}\bar{\sigma}}\phi^{\bar{\theta}}_\alpha\phi^{\bar{\sigma}}_\delta
\,,\end{equation} where the $(\rho_E)^i_\alpha,
\mathcal{C}^\alpha_{\beta\gamma}$ are the structure functions on $E$
and the $(\rho_{E'})^{\bar{i}}_{\bar{\alpha}}, {\bar{\mathcal{C}}}^{\bar\alpha}_{\bar\beta\bar\gamma}$ are the
structure functions on $E'$.

For more definitions and properties about the concept of Lie algebroid morphism, see for instance \cite{CLMMM-2006,HM-1990,Mart-2004,Mart-2005}.
\subsection{The prolongation of a Lie algebroid over a fibration }\label{prolong}

In this subsection we recall a particular kind of Lie algebroid that will be used later  (see \cite{CLMMM-2006,HM-1990,LMM-2005,Mart-2001}, for more details).

If
$(E,\lcf\cdot , \cdot\rcf_E,\rho_E)$ is a Lie algebroid over a manifold $Q$
and $\pi:P\to Q$ is a fibration, then $$\widetilde{\tau}_P\colon \mathcal{T}^EP= \displaystyle\bigcup_{{  p}\in
P}\mathcal{T}^E_{  p}P\to P,$$  where
\[\mathcal{T}^E_{  p}P=\{( {e}, {v}_{  {p}})\in E_{\pi(p)}\times T_{p}P\, |\,
\rho_E( {e})=T_{{  p}}\pi( {v}_{  {p}})\}\;\] is a Lie algebroid called the prolongation of the Lie algebroid $(E,\lcf\cdot , \cdot\rcf_E,\rho_E)$ (see for instance \cite{HM-1990,LMM-2005}). The anchor map of this Lie algebroid is  $\rho^\pi:\mathcal{T}^EP\to TP,\;\rho^\pi( {e}, {v}_{  p})= {v}_{  p}$. In this paper we consider two particular  Lie algebroid prolongations, one with $P=\rk\times (E\oplus \stackrel{k}{\ldots} \oplus E)$ and the other with $P=\rk\times (E^*\oplus \stackrel{k}{\ldots} \oplus E^*)$ (for more details see \cite{CLMMM-2006,HM-1990,LMM-2005,Mart-2001}).

If  $(q^i,u^\ell)$ are local coordinates on $P$ and  $\{e_\alpha\}$ is a local basis of sections of $E$, then a local basis of  $\widetilde{\tau}_P\colon \mathcal{T}^EP\to P$ is given by the family $\{\mathcal{X}_\alpha,\mathcal{V}_\ell\}$ where
\begin{equation}\label{base k-prol}
  \mathcal{X}_\alpha( {p}) =
(e_\alpha(\pi( {p}));\rho^i_\alpha(\pi( {p}))\displaystyle\frac{\partial
}{\partial q^i}\Big\vert_ {p}) \quad \makebox{and}\quad
 \mathcal{V}_\ell( {p})
=({ {0}}_{\pi( {p})};\displaystyle\frac{\partial}{\partial
u^\ell}\Big\vert_ {p})\,.
\end{equation}

The Lie bracket of two sections of $\mathcal{T}^EP$ is characterized by
 the relations
\begin{equation}\label{lie brack k-prol}\begin{array}{lll}
\lcf\mathcal{X}_\alpha,\mathcal{X}_\beta\rcf^{\pi}=
\mathcal{C}^\gamma_{\alpha\beta}\mathcal{X}_\gamma &
\lcf\mathcal{X}_\alpha,\mathcal{V}_\ell\rcf^{\pi}=0 &
\lcf\mathcal{V}_\ell,\mathcal{V}_\varphi\rcf^{\pi}=0\,,
\end{array}\end{equation}
and the exterior differential is therefore determined by
\begin{equation}\label{dtp}\begin{array}{lclclcl} \d^{\mathcal{T}^EP}q^i &=&
\rho^i_\alpha\mathcal{X}^\alpha\,,&\qquad & \d^{\mathcal{T}^EP}u^\ell
&=&
\mathcal{V}^\ell\\
\d^{\mathcal{T}^EP}\mathcal{X}^\gamma &=&
-\displaystyle\frac{1}{2}\mathcal{C}^\gamma_{\alpha\beta}
\mathcal{X}^\alpha\wedge\mathcal{X}^\beta\;,&\qquad &
\d^{\mathcal{T}^EP}\mathcal{V}^\ell&=&0\end{array}\end{equation}
where $\{\mathcal{X}^\alpha,\mathcal{V}^\ell\}$ is the dual basis of
$\{\mathcal{X}_\alpha,\mathcal{V}_\ell\}$.

\section{Classical Field Theories on Lie algebroids: a  $k$-cosymplectic
approach}\label{cftla}

In this section, the $k$-cosymplectic formalism for first order classical field theories (see \cite{LMORS-1998,LMS-2001}) is
extended to the general setting of Lie algebroids. Considering  a Lie
algebroid $E$ as a generalization of the tangent bundle $TQ$ of $Q$, we will define
the analog of the classical  field equations and their solutions,
and we study the analogs of the geometric structures of the standard
$k$-cosymplectic formalism. Lagrangian and Hamiltonian
  formalisms are developed in subsections
\ref{CFTLA} and \ref{Ha}, respectively, and it is verified that the
standard Lagrangian and Hamiltonian $k$-cosymplectic formalisms are
particular examples of the formalism  developed here. Throughout this section
we consider a Lie algebroid $(E,\lcf\cdot,\cdot\rcf_E,\rho_E)$ ($E$ for simplicity)  on the
manifold $Q$.

\subsection{Lagrangian formalism}\label{CFTLA}
First,  we will introduce some
  geometric ingredients which are necessary to develop the Lagrangian $k$-cosymplectic formalism on Lie algebroids.

\subsubsection{The manifold $\rk\times\stackrel{k}{\oplus} E$.}

The standard $k$-cosymplectic Lagrangian formalism is developed on the  bundle $\rktkq$, where $\tkq\equiv TQ\oplus\stackrel{k}{\ldots}\oplus TQ$ is the Whitney sum of $k$ copies of  $TQ$. Since we are thinking of a Lie algebroid $E$ as a substitute of the tangent bundle, it is natural to consider
$$\rk\times\ke\equiv\rk\times (E\oplus\stackrel{k}{\ldots}\oplus E)\,,$$  and the projection map $\widetilde{\mathrm{p}}\colon \rke\to Q$, given by $\widetilde{\mathrm{p}}(t^1,\ldots, t^k,{e_1}_{q},\ldots, {e_k}_{q})=q$.

Let us observe that the elements of $\rke$ have the following form:
$$(\mathbf{t},\mathbf{e}_{q})=(t^1,\ldots, t^k,{e_1}_{q},\ldots, {e_k}_{q})\,.$$

 If  $(q^i,y^\alpha)$ are local coordinates on
$\tau^{\,-1}(U)\subseteq E$, then the induced local coordinates $(t^A, q^i,y^\alpha_A)$ on $\widetilde{p}^{\,-1}(U)\subseteq\rke$ are given by
\begin{equation}\label{cke}
t^A(\mathbf{t},\mathbf{e}_{q})=t^A(\mathbf{t}),\quad
q^i(\mathbf{t},\mathbf{e}_{q})=q^i(q),\quad
y^\alpha_A(\mathbf{t},\mathbf{e}_{q})=y^\alpha({e_A}_{q})\,. \end{equation}
\subsubsection{The Lagrangian prolongation}\label{lagran prolong}

Consider the prolongation  of a Lie algebroid $E$ over the fibration $\widetilde{\mathrm{p}}\colon \rk\times\ke\to Q$, (see section \ref{prolong}),
\begin{equation}\label{lagrangian prolongation}\tec=\{(a_{q},v_{(\mathbf{t},\mathbf{e}_{q})})\in E\times T(\rke) /\;
\rho_E(a_{q})=T\widetilde{\mathrm{p}}(v_{(\mathbf{t},\mathbf{e}_{q})})\}\,,\end{equation}where $(\mathbf{t},\mathbf{e}_{q})\in\rke$. We deduce the following properties (see \cite{CLMMM-2006,LMM-2005,Mart-2001} for standard properties of the prolongation Lie algebroid):
\begin{enumerate}
\item $\tec\subset E\times T(\rke)$, with projection $$\widetilde{\tau}_{\rke}\colon  \tec\longrightarrow \rke$$ has a Lie algebroid structure
$(\lcf\cdot,\cdot\rcf^{\widetilde{\mathrm{p}}},\rho^{\widetilde{\mathrm{p}}}\,)$,
where the anchor map
$$\rho^{\widetilde{\mathrm{p}}}\colon  \tec\to
T(\rke)$$ is the canonical projection on the second factor. In the sequel, this induced Lie algebroid srtructure will be called the {\it Lagrangian prolongation}.

\item  If $(t^A,q^i,y^\alpha_A)$ are local coordinates  on $\rke$, then  the induced local coordinates on $\tec$ are $$(t^A,q^i,y^\alpha_A,z^\alpha,v_A,w^\alpha_ A)_{1\leq i\leq n,\;1 \leq A\leq k,\; 1\leq \alpha\leq m}$$
where
\begin{equation}\label{local coord cprol h}\begin{array}{lcllcl}
t^A(a_{q},v_{(\mathbf{t},\mathbf{e}_{q})}) &=& t^A(\mathbf{t})
  \;,\quad & z^\alpha(a_{q},v_{(\mathbf{t},\mathbf{e}_{q})})&=&y^\alpha(a_{q})
\;,\\\noalign{\medskip}
 q^i(a_{q},v_{(\mathbf{t},\mathbf{e}_{q})}) &=& q^i({q})
  \;,\quad &
  v_A(a_{q},v_{(\mathbf{t},\mathbf{e}_{q})})&=&v_{(\mathbf{t},\mathbf{e}_{q})}(t^A)\;,\\\noalign{\medskip}
y^\alpha_A(a_{q},v_{(\mathbf{t},\mathbf{e}_{q})})&=&y^\alpha_A(\mathbf{e}_{q})\;,
\quad & w^\alpha_A(a_{q},v_{(\mathbf{t},\mathbf{e}_{q})})
&=&v_{(\mathbf{t},\mathbf{e}_{q})}(y^\alpha_A)\;. \\
\end{array}
\end{equation}

\item The set
$\{\mathcal{Y}_A,\mathcal{X}_\alpha,\mathcal{V}^A_\alpha\}$ given by
\[\mathcal{Y}_A,\,\mathcal{X}_\alpha,\,\mathcal{V}^A_\alpha\colon  \rke\to\mathcal{T}^E(\rke)\]
{\small\begin{equation}\label{basec}
\mathcal{Y}_A(\mathbf{t},\mathbf{e}_{q})
=(0_{q};\displaystyle\frac{\partial}{\partial
t^A}\Big\vert_{(\mathbf{t},\mathbf{e}_{q})})\quad , \quad \mathcal{X}_\alpha(\mathbf{t},\mathbf{e}_{q}) =
(e_\alpha({q});\rho^i_\alpha({q})\displaystyle\frac{\partial
}{\partial q^i}\Big\vert_{(\mathbf{t},\mathbf{e}_{q})})\quad,\quad \mathcal{V}^A_\alpha(\mathbf{t},\mathbf{e}_{q})
=(0_{q};\displaystyle\frac{\partial}{\partial
y^\alpha_A}\Big\vert_{(\mathbf{t},\mathbf{e}_{q})})\end{equation}}
 is a local basis of $\sec{\tec}$, the set of sections of $\widetilde{\tau}_{\rke}$  (see \ref{base k-prol}).

\item   The anchor map $\rho^{\widetilde{\mathrm{p}}}\colon \mathcal{T}^E(\rke)\to T(\rke)$ allows us to associate a vector field with each section $\xi\colon \rke\to\tec$. Locally, if
        $$\xi=\xi^A\mathcal{Y}_A+\xi^\alpha\mathcal{X}_\alpha+\xi^\alpha_A\mathcal{V}^A_\alpha\in \sec{\tec}$$ then the associated vector field is given by
        \begin{equation}\label{rholksimc}
\rho^{\widetilde{\mathrm{p}}}(\xi)=\xi^A\derpar{}{t^A}+\rho^i_\alpha
\xi^\alpha\derpar{}{q^i} + \xi^\alpha_A\derpar{}{y^\alpha_A}\in
\vf(\rke)\,.\end{equation}

\item The Lie bracket of two sections of $\widetilde{\tau}_{\tec}$ is characterized by (see \ref{lie brack k-prol}):
\begin{equation}\label{lie brack tecL}\begin{array}{lclclclclcl}
\lcf\mathcal{Y}_A,\mathcal{Y}_B\rcf^{\widetilde{\mathrm{p}}}&=& 0\;,
&\quad &
\lcf\mathcal{Y}_A,\mathcal{X}_\alpha\rcf^{\widetilde{\mathrm{p}}}&=&
0\;, &\quad
&\lcf\mathcal{Y}_A,\mathcal{V}^B_\beta\rcf^{\widetilde{\mathrm{p}}}&=&
0\;,
\\\noalign{\medskip}
\lcf\mathcal{X}_\alpha,\mathcal{X}_\beta\rcf^{\widetilde{\mathrm{p}}}&=&
\mathcal{C}^\gamma_{\alpha\beta}\mathcal{X}_\gamma\;, &\quad &
\lcf\mathcal{X}_\alpha,\mathcal{V}^B_\beta\rcf^{\widetilde{\mathrm{p}}}&=&0\;,
&\quad &
\lcf\mathcal{V}^A_\alpha,\mathcal{V}^B_\beta\rcf^{\widetilde{\mathrm{p}}}&=&0\,.
\end{array}\end{equation}

\item If $\{\mathcal{Y}^A,\mathcal{X}^\alpha,
\mathcal{V}^\alpha_A\}$ is the dual basis of  $\{\mathcal{Y}_A,\mathcal{X}_\alpha, \mathcal{V}_\alpha^A\}$, then the exterior differential if given locally (see \ref{difEloc}) by
\begin{equation}\label{difksimc}
\begin{array}{l}
\d^{\tec}f=\derpar{f}{t^A}\mathcal{Y}^A+\rho^i_\alpha\derpar{f}{q^i}\mathcal{X}^\alpha
+ \derpar{f}{y^\alpha_A}\mathcal{V}^\alpha_A\,,\quad \makebox{for all } f\in \mathcal{C}^\infty(\rke)\\\noalign{\medskip} \d^{\tec}\mathcal{Y}^A =0\,,\quad
\d^{\tec}\mathcal{X}^\gamma =
-\displaystyle\frac{1}{2}\mathcal{C}^\gamma_{\alpha\beta}\mathcal{X}^\alpha\wedge\mathcal{X}^\beta\,,\quad
\d^{\tec}\mathcal{V}^\gamma_A =0\,.\end{array}
\end{equation}
\end{enumerate}

\begin{remark}\label{remark equivch}
In the particular case $E= TQ$, the manifold $\mathcal{T}^E(\rke)$ reduces to $T(\rktkq)$ since
\begin{equation}\label{equivalence}\hspace{-1cm}\begin{array}{lcl}&&\mathcal{T}^{TQ}(\rk\times\stackrel{k}{\oplus}TQ)=\mathcal{T}^{TQ}(\rktkq)
\\\noalign{\medskip} &=&\{(u_{q},v_{(\mathbf{t},\mathbf{w}_{q})})\in TQ\times
T(\rktkq)/ u_{q}=T\mathrm{p}_Q(v_{(\mathbf{t},\mathbf{w}_{q})})\}\\\noalign{\medskip}
&=&\{(T\mathrm{p}_Q(v_{(\mathbf{t},\mathbf{w}_{q})}),v_{(\mathbf{t},\mathbf{w}_{q})})\in
TQ\times T(\rktkq)/\; (\mathbf{t},\mathbf{w}_{q})\in
\rktkq\}\\\noalign{\medskip}&\equiv&\{v_{(\mathbf{t},\mathbf{w}_{q})}\in
T(\rktkq)/\; (\mathbf{t},\mathbf{w}_{q})\in \rktkq\}\equiv T(\rktkq)\;.
\end{array}\end{equation}  \end{remark}

\subsubsection{The Liouville sections and  vertical endomorphisms}

On $\tec$ we define two families of canonical objects, {\it Liouville sections} and {\it vertical endomorphism} which correspond to the {\it Liouville vector fields} and {\it canonical tensor fields} on $\rktkq$ (see \cite{LMS-2001,MSV-2005,MSV-2009}.)

\paragraph{\bf $A{th}$-vertical lifting (see for instance \cite{CLMMM-2006}).}   An element $(a_{q},v_{(\mathbf{t},\mathbf{e}_{q})})$
of $\mathcal{T}^E(\rke)$
is said to be vertical if
$$\widetilde{\tau}_1(a_{q},v_{(\mathbf{t},\mathbf{e}_{q})})=0_{q}\in
E\,,$$ where
$$\begin{array}{rrcl}
\widetilde{\tau}_1:&\mathcal{T}^E(\rke)&\to& E,\\\noalign{\medskip}
&(a_{q},v_{(\mathbf{t},\mathbf{e}_{q})}) & \mapsto &\;
a_{q}\end{array}$$ is the projection on the first factor $E$. The vertical elements of $
\mathcal{T}^E(\rke)$ are thus of the form
$$(0_{q},v_{(\mathbf{t},\mathbf{e}_{q})})\in \mathcal{T}^E(\rke)$$
where $v_{(\mathbf{t},\mathbf{e}_{q})}\in T(\rke)$ and
$(\mathbf{t},\mathbf{e}_{q})\in \rke$. In particular, the tangent vector $v_{(\mathbf{t},\mathbf{e}_{q})}$ is $\widetilde{\mathrm{p}}$-vertical, since by  (\ref{lagrangian prolongation})
$$0=T_{(\mathbf{t},\mathbf{e}_{q})}\widetilde{\mathrm{p}}\big(v_{(\mathbf{t},\mathbf{e}_{q})}\big)\,.$$

In a local coordinate system  $(t^A,q^i,y^\alpha_A)$  on $\rke$, if $(a_{q},v_{(\mathbf{t},\mathbf{e}_{q})})\in\tec$ is vertical, then $a_{q}=0_q$ and
$$v_{(\mathbf{t},\mathbf{e}_{q})}=u_A\derpar{}{t^A}\Big\vert_{(\mathbf{t},\mathbf{e}_{q})}+  u^\alpha_B \displaystyle\frac{\partial }{\partial
y^\alpha_B}\Big\vert_{(\mathbf{t},\mathbf{e}_{q})}\in
T_{(\mathbf{t},\mathbf{e}_{q})}( \rke)\,.$$

\begin{definition}\label{lvastkec}

For each $A=1,\ldots, k$, the vertical  $A{th}$-lifting is defined as the mapping
 \begin{equation}\label{a-levantamientoc}
\begin{array}{rcl}
{\cdot}^{{\mathbf v}_A}:E\times_Q(\rke) & \longrightarrow &
\mathcal{T}^E(\rke)
\\\noalign{\medskip}
 (a_q,\mathbf{t},\mathbf{e}_{q}) & \longmapsto &
(a_q,\mathbf{t},\mathbf{e}_{q})^{{\mathbf v}_A}=\left(0_{q},(a_q)^{v_A}_{(\mathbf{t},\mathbf{e}_{q})}\right) \\
\end{array}
\end{equation} where $a_q\in E,\; (\mathbf{t},\mathbf{e}_{q})=( t^1,\ldots, t^k ,{e_1}_{q},\ldots, {e_k}_{q})\in \rke$ and the vector $\;(a_q)^{v_A}_{(\mathbf{t},\mathbf{e}_{q})}\in
T_{(\mathbf{t},\mathbf{e}_{q})} (\rke)$ is given by
\begin{equation}\label{verticalc}
(a_q)^{v_A}_{(\mathbf{t},\mathbf{e}_{q})}f=\displaystyle\frac{d}{ds}\Big\vert_{s=0}f(\mathbf{t},
e_{1_{q}},\ldots, e_{A_{q}}+s a_{q},\ldots,
e_{k_{q}})\;,\quad 1\leq A\leq k\;,
\end{equation} for an arbitrary function $f\in \mathcal{C}^\infty(\rke)$.
\end{definition}

The local expression of
$(a_q)^{v_A}_{(\mathbf{t},\mathbf{e}_{q})}$ is
\begin{equation}\label{localvertc}
(a_q)^{v_A}_{(\mathbf{t},\mathbf{e}_{q})}=y^\alpha(a_q)\displaystyle\frac{\partial}{\partial
y^\alpha_A}\Big\vert_{(\mathbf{t},\mathbf{e}_{q})}\in
T_{(\mathbf{t},\mathbf{e}_{q})}( \rke)\;,\quad 1\leq A\leq k\;.
\end{equation}

Since $(a_q)^{v_A}_{(\mathbf{t},\mathbf{e}_{q})}\in T_{(\mathbf{t},\mathbf{e}_{q})}(\rke)$ is $\widetilde{\mathrm{p}}$-vertical, and from
 (\ref{basec}),  (\ref{a-levantamientoc}) and (\ref{localvertc}) we deduce that locally
\begin{equation}\label{localxiac}\hspace{-0.5cm}
\begin{array}{lcl}
(a_q,\mathbf{t},\mathbf{e}_{q})^{{\mathbf v}_A}&=&(0_{q},y^\alpha(a_q)\displaystyle\frac{\partial}{\partial
y^\alpha_A}\Big\vert_{(\mathbf{t},\mathbf{e}_{q})}) =
y^\alpha(a_q)\mathcal{V}^A_\alpha(\mathbf{t},\mathbf{e}_{q})\,,\end{array}\quad
1\leq A\leq k\;.\end{equation}

\paragraph{\bf Vertical endomorphisms on $\mathcal{T}^E(\rke)$.}

One of the most important family of canonical geometric elements on $\tec$ is the family of vertical endomorphisms $\widetilde{S}^1,\ldots, \widetilde{S}^k$. This family plays the role of the canonical tensor fields $S^1,\ldots, S^k$ in the standard case (see, for instance \cite{LMS-2001,MSV-2005,MSV-2009,RRSV-07,RRSV-09}).

\begin{definition}\label{endvertA}
For  $A=1,\ldots, k$ the {\rm $A{th}$-vertical endomorphism} on  $\mathcal{T}^E(\rke)$ is the mapping
\begin{equation}\label{jtildeA}\begin{array}{rccl}
\widetilde{S}^A:&\mathcal{T}^E(\rke) & \to &
\mathcal{T}^E(\rke)\\\noalign{\medskip}
&(a_{q},v_{(\mathbf{t},\mathbf{e}_{q})})&\mapsto
&\widetilde{S}^A(a_{q},v_{(\mathbf{t},\mathbf{e}_{q})})=
(a_{q},\mathbf{t},\mathbf{e}_{q})^{{\mathbf v}_A}\,,
\end{array}\end{equation} where
$a_{q}\in E,\;(\mathbf{t},\mathbf{e}_{q})\in\rke$ and
$v_{(\mathbf{t},\mathbf{e}_{q})}\in T_{(\mathbf{t},\mathbf{e}_{q})}(\rke)$.
\end{definition}

Locally,  let $\{\mathcal{Y}_A,\mathcal{X}_\alpha,\mathcal{V}^A_\alpha\}$ be a local basis of $\sec{\tec}$ and let $\{\mathcal{Y}^A,\mathcal{X}^\alpha,\mathcal{V}_A^\alpha\}$ be its dual basis.  The corresponding local expression of $\widetilde{S}^A$ is
%From (\ref{basec}) and (\ref{localxiac}) one deduces that $\widetilde{S}^A$ has the following local expression:
\begin{equation}\label{localtildeSAksim}
\widetilde{S}^A=\displaystyle\sum_{\alpha=1}^m\mathcal{V}^A_\alpha\otimes\mathcal{X}^\alpha\;,
\quad 1\leq A\leq k\;.
\end{equation}

\begin{remark}\label{vertical tensor}\
\begin{enumerate}
\item In the standard case ($E=TQ,\, \rho=id_{TQ}$), the $\widetilde{S}^A$ constitutes the canonical tensor fields $S^1,\ldots, S^k$ on $\rktkq$ ( see, for instance, \cite{LMS-2001,MSV-2005,MSV-2009,RRSV-07,RRSV-09}).
\item The endomorphisms $\widetilde{S}^1,\ldots,\widetilde{S}^k$ defined here allows us to introduce the {\it Lagrangian sections} when we develop the  $k$-cosymplectic Lagrangian formalism on Lie algebroids. Moreover these mappings give a characterization of certain sections  of  $\tec$ which we consider in the following subsection.
\end{enumerate}
 \end{remark}

 \paragraph{\bf The Liouville sections}\label{Lagform1}
The  {\rm $A{th}$ Liouville section} $\widehat{\Delta}_A$ is the section of
$\widetilde{\tau}_{\rke}:\mathcal{T}^E(\rke)\to \rke$ given by
$$
\begin{array}{rcl}
\widehat{\Delta}_A: \rk\times\stackrel{k}{\oplus} E & \to &
\mathcal{T}^E(\rke)\\\noalign{\medskip}
(\mathbf{t},\mathbf{e}_{q})&\mapsto
&\widehat{\Delta}_A(\mathbf{t},\mathbf{e}_{q})=
(pr_A(\mathbf{t},\mathbf{e}_{q}), \mathbf{t} ,\mathbf{e}_{q})^{{\mathbf v}_A}=({e_A}_{q}, \mathbf{t} ,\mathbf{e}_{q})^{{\mathbf v}_A}
\end{array}\;,
$$ where  $pr_A:\rke\to E$ is the canonical projection of $\rke$ over the $A{th}$ copy of
  $E$.  From the local expression (\ref{localxiac}) of ${\cdot}^{V_A}$, and since
$$y^\alpha({e_A}_{q})=y^\alpha_A(\mathbf{t},{e_1}_{q},\ldots,
{e_k}_{q})=y^\alpha_ A(\mathbf{t},\mathbf{e}_{q}),$$  $\widehat{\Delta}_A$ has the local expression
\begin{equation}\label{Liouville kcosim}
\widehat{\Delta}_A=\displaystyle\sum_{\alpha=1}^m
y^\alpha_A\mathcal{V}^A_\alpha\;, \quad 1\leq A\leq k\;.
\end{equation}
\begin{remark} \label{liouvilleA kcsim}
In the standard case,  $\widehat{\Delta}_A$ is the $A{th}$-Liouville vector field $\Delta_A$ on  $\rktkq$, (see for instance \cite{LMS-2001,MSV-2005,RRSV-07,RRSV-09}).
\end{remark}

In the standard Lagrangian $k$-cosymplectic formalism, the Liouville vector fields $\Delta_1,\ldots, \Delta_k$ allows us to define the energy function. Analogously as we will see below, the energy function can be defined in the Lie algebroid setting using the Liouville sections $\widetilde{\Delta}_1,\ldots, \widetilde{\Delta}_k$.

\subsubsection{Second order partial differential equations ({\sc sopde's}).}
\label{Sec 8.1.2.}

In the standard $k$-cosymplectic Lagrangian formalism one obtains the solutions of the Euler-Lagrange equations  as integral sections of certain second-order partial differential equations ({\sc sopde} in the sequel) on $\rktkq$. In order to introduce the analogous object on Lie algebroids, we note that in the standard case a {\sc sopde} $\xi$ is a $k$-vector field on $\rktkq$, that is, a section of
$$T^1_k(\rktkq)\equiv T(\rktkq)\oplus\stackrel{k}{\ldots}\oplus T(\rktkq)\to \rktkq\,,$$ which satisfies certain properties. Since $T^1_k(\rktkq)$ is the Whitney sum of $k$ copies of $T(\rktkq)$, it is natural to think that, in the Lie algebroid context, the appropriate space would be    the Whitney sum of $k$ copies of $\mathcal{T}^E(\rke)$, that is
$$\tkec=\tec\oplus\stackrel{k}{\ldots}\oplus\tec\,.$$ We denote by $\widetilde{\tau}^k_{\rke}$ its canonical projection on $\rke$.

\begin{definition} A {\rm second order partial differential equation ({\sc sopde})} on $\rke$ is a map
$\,\xi=(\xi_1,\ldots,\xi_k):\rk\times\stackrel{k}{\oplus} E\to
\tkec $ which is  a section of $\widetilde{\tau}^k_{\rke}$ and satisfies the equations
$$
\widetilde{S}^A(\xi_A)=\widehat{\Delta}_A \quad \makebox{and} \quad
\mathcal{Y}^B(\xi_A)=\delta^B_A\,,\qquad 1\leq A,B\leq k\,.
$$
\end{definition}

Since $\tkec$ is the Whitney sum of $k$ copies of $\tec$, we deduce that to give a section $\xi$ of $\widetilde{\tau}^k_{\rke}$ is equivalent to giving a family of $k$ sections $\xi_1,\ldots, \xi_k$ of the Lagrangian prolongation $\mathcal{T}^E(\rke)$ obtained by projection $\xi$ on each factor.

One easily deduces that the local expression of a {\sc sopde} $\xi=(\xi_1,\ldots, \xi_k)$ is
\begin{equation}\label{sopde c}\xi_A=\mathcal{Y}_A+
y_A^\alpha\mathcal{X}_\alpha +
(\xi_A)_B^\alpha\mathcal{V}_\alpha^B\;,\end{equation} where $(\xi_A)^B_\alpha\in\mathcal{C}^\infty(\rke)$.

\begin{lemma} Let $\xi=(\xi_1,\ldots,\xi_k)\colon  \rke\to \tkec$ be a
section of $\widetilde{\tau}^k_{\rke}$. Then
$$(\rho^{\widetilde{\mathrm{p}}}(\xi_1),\ldots,\rho^{\widetilde{\mathrm{p}}}(\xi_k))\colon
\rke\to T^1_k(\rke)$$ is a $k$-vector field on $\rke$, where $\rho^{\widetilde{\mathrm{p}}}\colon  \tec\equiv
E\times_{TQ}T(\rke)\to T(\rke)$ is the anchor map of the Lie algebroid $\tec$.
\end{lemma}

\proof Directly by section \ref{lagran prolong} {\it (6)}. \qed

In local coordinates
\begin{equation}\label{sopde asso co}\rho^{\widetilde{\mathrm{p}}}(\xi_A)=
\derpar{}{t^A}+\rho^i_\alpha y^\alpha_A\derpar{}{q^i} +
(\xi_A)^\alpha_B\derpar{}{y^\alpha_B}\in\vf(\rke)\,.\end{equation}

\begin{definition}
A map
$\eta\colon  U\subseteq\rk\to \rke$ is an integral section of a
{\sc sopde} $\xi=(\xi_1,\ldots,\xi_k)$ if $\eta$ is an integral section of the $k$-vector field
$(\rho^{\widetilde{\mathrm{p}}}(\xi_1),\ldots,\rho^{\widetilde{\mathrm{p}}}(\xi_k))$
associated with $\xi$, that is,
\begin{equation}\label{int sect co}
\rho^{\widetilde{\mathrm{p}}}(\xi_A)(\eta(\mathbf{t}))=
\eta_*(\mathbf{t})\left(\displaystyle\frac{\partial}{\partial t^A}\Big\vert_{
\mathbf{t}}\right)\;,\quad 1\leq A\leq k\;.
\end{equation}\end{definition}

In $\eta$ is written locally as $\eta(\mathbf{t})=(\eta_A(\mathbf{t}),\eta^i(\mathbf{t}),\eta^\alpha_A(\mathbf{t})),$ then from  (\ref{sopde asso co}) we deduce that (\ref{int sect co}) is locally equivalent to the identities
\begin{equation}\label{integral sect}
\displaystyle\frac{\partial \eta_B}{\partial
t^A}\Big\vert_{\mathbf{t}}=\delta^A_B \;,\quad \displaystyle\frac{\partial
\eta^i}{\partial t^A}\Big\vert_{\mathbf{t}}=
\eta^\alpha_A(\mathbf{t})\rho^i_\alpha\;,\quad \displaystyle\frac{\partial
\eta^\beta_B}{\partial
t^A}\Big\vert_{\mathbf{t}}=(\xi_A)^\beta_B(\eta(\mathbf{t}))\;.
\end{equation}

\subsubsection{Lagrangian formalism.}\label{Sec 8.1.3.}

In this  section we develop an intrinsic and global geometric framework that allows  us to write the Euler-Lagrange equations associated with a Lagrangian function $L\colon \rke\to\r$ on a Lie algebroid. We first introduce some geometric elements associated with $L$.

\paragraph{\bf Poincar\'{e}-Cartan or Lagrangian sections} {\it The Poincar\'{e}-Cartan} $1$-sections $\Theta_L^A$ are defined by
$$
\begin{array}{rcc}
\Theta_L^A: \rke & \longrightarrow & (\mathcal{T}^E(\rke))^{\;*}
\\\noalign{\medskip}
(\mathbf{t},\mathbf{e}_{q}) & \longmapsto &
\Theta_L^A(\mathbf{t},\mathbf{e}_{q})\end{array}
$$
where $\Theta_L^A(\mathbf{t},\mathbf{e}_{q})\colon (\mathcal{T}^E(\rke) )_{(\mathbf{t},\mathbf{e}_{q})} \to \r$ is the linear mapping
 defined by
\begin{equation}\label{ThetaLAc}
(\Theta_L^A)(\mathbf{t},\mathbf{e}_{q})(a_{q},v_{(\mathbf{t},\mathbf{e}_{q})})=(\d^{\tec}
L)_{(\mathbf{t},\mathbf{e}_{q})}((\widetilde{S}^A)_{(\mathbf{t},\mathbf{e}_{q})}(a_{q},v_{(\mathbf{t},\mathbf{e}_{q})}))
\;.
\end{equation}

Using (\ref{difksimc}) with
$f=L$, \begin{equation}\label{theta c
al}\begin{array}{lcl}
(\Theta_L^A)(\mathbf{t},\mathbf{e}_{q})(a_{q},v_{(\mathbf{t},\mathbf{e}_{q})})&=&(\d^{\tec}
L)_{(\mathbf{t},\mathbf{e}_{q})}((\widetilde{S}^A)_{(\mathbf{t},\mathbf{e}_{q})}(a_{q},v_{(\mathbf{t},\mathbf{e}_{q})}))\\\noalign{\medskip}
&=&
[\rho^{\widetilde{\mathrm{p}}}((\widetilde{S}^A)_{(\mathbf{t},\mathbf{e}_{q})}(a_{q},v_{(\mathbf{t},\mathbf{e}_{q})}))]L\;,
\end{array}\end{equation} where $(\mathbf{t},\mathbf{e}_{q})\in \rke$,
$(a_{q},v_{(\mathbf{t},\mathbf{e}_{q})})\in
[\mathcal{T}^E(\rke)]_{(\mathbf{t},\mathbf{e}_{q})}$ and
$$\rho^{\widetilde{\mathrm{p}}}((\widetilde{S}^A)_{(\mathbf{t},\mathbf{e}_{q})}(a_{q},v_{(\mathbf{t},\mathbf{e}_{q})}))\in
T_{(\mathbf{t},\mathbf{e}_{q})}(\rke).$$

The Poincar\'{e}-Cartan $2$-sections
$$\Omega_L^A:\rke \to
\Lambda^2(\mathcal{T}^E(\rke))^{\;*},\;1\leq
A\leq k$$ are defined by
$$
\Omega_L^A\colon  =-\d^{\tec}\Theta_L^A\;,\quad 1\leq A\leq k\;.
$$

To find the local expression of $\Theta_L^A$ and $\Omega_L^A$, consider  $\{\mathcal{Y}_B,\mathcal{X}_\alpha,\;\mathcal{V}^B_\alpha\} $, a local basis of $\sec{\mathcal{T}^E(\rke)}$,  and its dual basis
$\{\mathcal{Y}^B,\mathcal{X}^\alpha,\;\mathcal{V}_B^\alpha\} $. From
(\ref{rholksimc}), (\ref{localtildeSAksim}) and (\ref{theta c al}), we deduce that
\begin{equation}\label{local thetac}
\Theta_L^A=\displaystyle\frac{\partial L}{\partial y^\alpha
_A}\mathcal{X}^\alpha \;,\qquad 1\leq A\leq k\;,
\end{equation}and from the local expressions (\ref{rholksimc}), (\ref{lie brack
tecL}), (\ref{difksimc})  and (\ref{local thetac}),
{\small\begin{equation}\label{local omega co}
\Omega_L^A = \displaystyle\frac{1}{2} \left(\rho^i_\beta \displaystyle\frac{\partial
^2 L}{\partial q^i\partial y^\alpha_A} - \rho^i_\alpha
\displaystyle\frac{\partial ^2 L}{\partial q^i\partial y^\beta_A} +
\mathcal{C}^\gamma_{\alpha\beta}\displaystyle\frac{\partial L}{\partial
y^\gamma_A}\right) \mathcal{X}^\alpha \wedge \mathcal{X}^\beta
+
\derpars{L}{t^B}{y^\alpha_A}\mathcal{X}^\alpha\wedge \mathcal{Y}^B +
\displaystyle\frac{\partial ^2 L}{\partial y^\beta_ B\partial y^\alpha_A}\,
\mathcal{X}^\alpha \wedge \mathcal{V}_B^\beta\,.
\end{equation}}

We say that the lagrangian $L$ is {\it regular} if  the matrix  $(\frac{\partial^2L}{\partial y^\alpha_A\partial y^\beta_B})$ is non-singular.

\begin{remark}  When we consider the particular case $E= TQ$ and $\rho=id_{TQ}$,
  $$\Omega_L^A(X,Y)=\omega_L^A(X,Y)\,, \qquad 1\leq A\leq
k\,,$$ where $X,Y$ are   vector fields on $\rktkq$ and
$\omega_L^1,\ldots, \omega_L ^{\,k}$ are the Lagrangian $2$-forms of the standard $k$-cosymplectic Lagrangian formalism, see for instance \cite{MSV-2005,RRSV-07,RRSV-09}. \end{remark}

\paragraph{\bf The energy function.} The {\it energy function}
   $
 E_L:\rke \to \r$
  defined by the Lagrangian $L$ is
$$
E_L=\displaystyle\sum_{A=1}^{\,k}\rho^{\widetilde{\mathrm{p}}}(\widehat{\Delta}_A)L-L\;.
$$ and from (\ref{rholksimc}) and (\ref{Liouville kcosim}) one deduces that $E_L$
is locally given  by
\begin{equation}\label{local enerco}
E_L=\displaystyle\sum_{A=1}^{\,k} y^\alpha_A\displaystyle\frac{\partial L}{\partial
y^\alpha_A}- L\in \mathcal{C}^\infty(\rke)\;.
\end{equation}

\paragraph{\bf Morphisms} We generalize the
 Euler-Lagrange equations and their solutions to the case of Lie algebroids in terms of   Lie algebroid morphisms.

  In the standard Lagrangian $k$-cosymplectic  formalism, a solution of the Euler-Lagrange
equations is a field $\phi:\rk\to Q$ with a first prolongation
$\phi^{[1]}:\rk\to \rk\times T^1_kQ$  satisfying those equations, that is,
$$
\displaystyle \sum_{A=1}^k\displaystyle\frac{\partial}{\partial
t^A}\Big\vert_{ {\mathbf{t}}} \left(\frac{\displaystyle\partial
L}{\displaystyle
\partial v^i_A}\Big\vert_{\phi^{[1]}( {\mathbf{t}})} \right)= \frac{\displaystyle \partial
L}{\displaystyle
\partial q^i}\Big\vert_{\phi^{[1]}( {\mathbf{t}})}\,.
$$

The map $\phi$ naturally induces  the Lie algebroid morphism
\[\xymatrix
{T\rk  \ar[r]^-{T\phi}\ar[d]_{\tau_{\rk}} & TQ\ar[d]^-{\tau_Q}\\
\rk\ar[r]_-{\phi} & Q}\] and in terms of the canonical basis of sections of $\tau_{\rk}$,
$\;\left\{\displaystyle\frac{\partial}{\partial t^1},\ldots,
\displaystyle\frac{\partial}{\partial t^k}\right\}$, the  first
prolongation of $\phi$, $\,\phi^{[1]}$, can be written as
\[\phi^{[1]}(\mathbf{t})=(\mathbf{t},T_\mathbf{t}\phi(\displaystyle\frac{\partial}{\partial
t^1}\Big\vert_{\mathbf{t}}),\ldots,T_\mathbf{t}\phi(\displaystyle\frac{\partial}{\partial
t^k}\Big\vert_{\mathbf{t}}))\,.\]

For a general Lie algebroid we shall derive Euler-Lagrange equations for field theories on Lie algebroids using as a main tool Lie  algebroid morphisms $\Phi=(\overline{\Phi},\underline{\Phi})$,
 \[\xymatrix
{T\rk  \ar[r]^-{\overline{\Phi}}\ar[d]_{\tau_{\rk}} & E\ar[d]^-{\tau}\\
\rk\ar[r]_-{\underline{\Phi}} & Q}\] with an associated map $\widetilde{\Phi}:\rk\to\rk\times\stackrel{k}{\oplus}
E$
\[\begin{array}{rcl}
\widetilde{\Phi}:\rk &\to &\rk\times
\stackrel{k}{\oplus}E\equiv\rk\times
E\oplus\stackrel{k}{\ldots}\oplus E\\\noalign{\medskip} \mathbf{t} &\to
&(\mathbf{t},\overline{\Phi}(e_1(\mathbf{t})),\ldots,
\overline{\Phi}(e_k(\mathbf{t})))\;.
\end{array}\] where
$\{e_A\}_{A=1}^k$   is a fixed local basis of local  sections of $T\rk$.

If $(t^A)$ and $(q^i)$ are local coordinate systems on $\rk$ and
$Q$, respectively; $\{e_A\}$ and  $\{e_\alpha\}$ local basis of sections of
$\tau_{\rk}$ and  $E$, respectively; and
  $\{e^A\}$ and $\{e^\alpha\}$ the respective dual bases; then
$\underline{\Phi}(\mathbf{t})=(\phi^i(\mathbf{t}))$ and $\Phi^* e^\alpha=\phi^\alpha_A
e^A$ for certain local functions $\phi^i$ and $\phi^\alpha_A$ on
$\rk$, the associated map $\widetilde{\Phi}$ is  given locally
by $\widetilde{\Phi}(\mathbf{t})=(t^A,\phi^i(\mathbf{t}),\phi^\alpha_A(\mathbf{t}))$, and the Lie algebroid morphism conditions (\ref{morp cond}) are \begin{equation}\label{morpcond}
  \rho^i_
 \alpha\phi^\alpha_A = \displaystyle\frac{\partial \phi^i}{\partial t^A} \quad , \quad
0=\displaystyle\frac{\partial \phi^\alpha_A}{\partial t^B} -
\displaystyle\frac{\partial \phi^\alpha_B}{\partial t^A}
+\mathcal{C}^\alpha_{\beta\gamma}\phi^\beta_B\phi^\alpha_A
\,.\end{equation}

\begin{remark} In the standard case ($E=TQ$),
the morphism conditions reduce to
\[\phi^i_A=\displaystyle\frac{\partial \phi^i}{\partial t^A} \quad
\makebox{and}\quad  \displaystyle\frac{\partial \phi^i_A}{\partial
t^B}=\displaystyle\frac{\partial \phi^i_B}{\partial t^A}\,,\] i.e., the standard first-order prolongations
 of fields $\phi:\rk\to Q$.\end{remark}

\paragraph{\bf The Euler-Lagrange equations.} Given a regular Lagrangian function $L\colon \rke\to \r\,,$ it is natural to consider sections
$\mathbf{\xi}_L=(\xi_1,\ldots,\xi_k)$ of $(\mathcal{T}^E)^1_k(\rke)=\tec\oplus\stackrel{k}{\ldots}\oplus\tec\to \rke$ such that
\begin{equation}\label{ec ge EL co}
 \mathcal{Y}^B(\xi_A)=\delta_A^B\quad,\quad
    \displaystyle\sum_{A=1}^k\imath_{\xi_A}\Omega_L^A =\d^{\tec} E_L+
\displaystyle\sum_{A=1}^k\frac{\partial L}{\partial t^A}\mathcal{Y}^A\;.
\end{equation}
equation (\ref{ec ge EL co}) being the analog of the geometric Euler-Lagrange equations of the standard $k$-cosymplectic Lagrangian formalism.

\begin{theorem}\label{algeform}
Let $L:\rk \times\stackrel{k}{\oplus}E\to\r$ be a regular Lagrangian, and
$\xi_1,\ldots, \xi_k$  $k$ sections of $\widetilde{\tau}_{\rke}\colon$ $ \tec\to \rke$  such that
\[
\mathcal{Y}^B(\xi_A)=\delta_A^B\quad,\quad
    \displaystyle\sum_{A=1}^k\imath_{\xi_A}\Omega_L^A =\d^{\tec} E_L+
\displaystyle\sum_{A=1}^k\frac{\partial L}{\partial t^A}\mathcal{Y}_A\;.\]
Then:
\begin{enumerate}
\item $\mathbf{\xi}_L=(\xi_1,\ldots, \xi_k)$ is a {\sc sopde}.
\item If $\widetilde{\Phi}:\rk\to \rk\times\stackrel{k}{\oplus}E\,$ is  the
map associated with a  Lie algebroid morphism between $T\rk$ and
$E$, and is an integral section of $\mathbf{\xi}_L$, then it
is a solution of the {\it  Euler-Lagrange   equations of field theories on Lie algebroids},
that is,
\begin{equation}\label{eq E-L cosim}\begin{array}{rcl}
\displaystyle\sum_{A=1}^k\displaystyle\frac{\partial}{\partial t^A}\left(
\displaystyle\frac{\partial  L}{\partial
y^\alpha_A}\Big\vert_{\widetilde{\Phi}(\mathbf{t})}\right) &=&
\rho^i_\alpha \displaystyle\frac{\partial L}{\partial
q^i}\Big\vert_{\widetilde{\Phi}(\mathbf{t})} -
\phi^\beta_A(\mathbf{t})\mathcal{C}^\gamma_{\alpha\beta}\displaystyle\frac{\partial
L}{\partial y^\gamma_A}\Big\vert_{\widetilde{\Phi}(\mathbf{t})}\;,
 \\\noalign{\medskip}
\displaystyle\frac{\partial \phi^i}{\partial t^A}\Big\vert_{\mathbf{t}} &=&
\phi^\alpha_A(\mathbf{t})\rho^i_\alpha
\\\noalign{\medskip}
0&=&\displaystyle\frac{\partial \phi^\alpha_A}{\partial t^B}\Big\vert_{\mathbf{t}} -
\displaystyle\frac{\partial \phi^\alpha_B}{\partial t^A}\Big\vert_{\mathbf{t}}
+\mathcal{C}^\alpha_{\beta\gamma}\phi^\beta_B(\mathbf{t})\phi^\gamma_A(\mathbf{t})\; .
\end{array}
\end{equation}
\end{enumerate}
\end{theorem}

\proof  The proof is analogous to the one in Theorem 4.18 in \cite{LMSV-09}.

In this case one obtains that if
$\xi_L=(\xi_1,\ldots,\xi_k):\rke\to\tkec$ is a solution to (\ref{ec ge EL co}) then:
\begin{enumerate}
\item $\xi_L$ is a {\sc sopde} on $\tec$. With respect to a local coordinate system $(t^A,q^i,y^\alpha_A)$ on $\rke$ and a local basis $\{e_\alpha\}$ of $\Sec{E}$ it is given locally by
$$\xi_A= \mathcal{Y}_A+y_A^\alpha
\mathcal{X}_\alpha + (\xi_A)^\alpha_B \mathcal{V}^B_\alpha $$  $(\xi_A)^\alpha_B$ being functions on $\rke$;
\item the functions $(\xi_A)^\alpha_B\in \mathcal{C}^\infty(\rke)$
satisfy the following equations:\end{enumerate}
\begin{equation}\label{eqla3}\derpars{L}{t^A}{y^\alpha_A}+y^\beta_A \rho^i_\beta\derpars{L}{q^i}{y^\alpha_A}+
(\xi_A)^\beta_B \derpars{L}{y^\beta_B}{y^\alpha_A} =
\rho^i_\alpha\derpar{L}{q^i} - y^\beta_A
\mathcal{C}^\gamma_{\alpha\beta}\derpar{L}{y^\gamma_A}
\,.\end{equation}

If the map $\widetilde{\Phi}:\rk\to\rk\times\stackrel{k}{\oplus}E$ associated with a Lie algebroid morphism $\Phi\colon T\rk\to E$ and defined by
$\widetilde{\Phi}(\mathbf{t})=(\mathbf{t},\phi^i(\mathbf{t}),
\phi^\alpha_A(\mathbf{t}))$,  is  an integral section of
$\xi_L$, then by condition (\ref{integral sect}) and equations (\ref{eqla3}) we obtain%
%
% be the associated map to a morphisms of Lie algebroids $\Phi=(\overline{\Phi},\underline{\Phi})$ between $\tau_{\rk}:T\rk\to\rk$ and $\tau:E\to Q$. If $\widetilde{\Phi}(\mathbf{t})=(\mathbf{t},\phi^i(\mathbf{t}),
%\phi^\alpha_A(\mathbf{t}))$ is an integral section of
%$\xi$, then from (\ref{integral sect}),
%(\ref{morpcondco}) and (\ref{eqla3}) we obtain that $\widetilde{\Phi}$ is a solution to
$$\begin{array}{rcl} \displaystyle\sum_{A=1}^k\displaystyle\frac{\partial}{\partial
t^A}\left( \displaystyle\frac{\partial  L}{\partial
y^\alpha_A}\Big\vert_{\widetilde{\Phi}(\mathbf{t})}\right) &=&
\rho^i_\alpha \displaystyle\frac{\partial L}{\partial
q^i}\Big\vert_{\widetilde{\Phi}(\mathbf{t})} -
\phi^\beta_A(\mathbf{t})\mathcal{C}^\gamma_{\alpha\beta}\displaystyle\frac{\partial
L}{\partial y^\gamma_A}\Big\vert_{\widetilde{\Phi}(\mathbf{t})}\;,
\\\noalign{\medskip}
\displaystyle\frac{\partial \phi^i}{\partial t^A}\Big\vert_{\mathbf{t}} &=&
\phi^\alpha_A(\mathbf{t})\rho^i_\alpha\;,
\\\noalign{\medskip}
0&=&\displaystyle\frac{\partial \phi^\alpha_A}{\partial
t^B}\Big\vert_{\mathbf{t}} - \displaystyle\frac{\partial
\phi^\alpha_B}{\partial t^A}\Big\vert_{\mathbf{t}}
+\mathcal{C}^\alpha_{\beta\gamma}\phi^\beta_B(\mathbf{t})\phi^\gamma_A(\mathbf{t})
\end{array}
$$ where the last two equations are consequence of the morphism conditions (\ref{morpcond}).\qed

If $E$ is the standard Lie algebroid $TQ$, the previous equations are  the classical Euler-Lagrange equations for the Lagrangian $L\colon\rk\times T^1_kQ\to \r$. In what follows (\ref{eq E-L cosim}) will be called the {\it Euler-Lagrange equations of field theories on Lie algebroids}.

\begin{remark}\
\begin{enumerate}
\item Equations (\ref{eq E-L cosim}) are obtained by E. Martinez \cite{Mart-2005} using a variational approach in the multisymplectic framework.

\item If $L$ does not depends on $\mathbf{t}$, then it can be considered as a map $L\colon \ke\to \r$. In this case the sections $\Omega_L^A$ can be thought as sections of $\mathcal{T}^E(\ke)$ and from (\ref{ec ge EL co}) we deduce the $k$-symplectic Euler-Lagrange equations  on Lie algebroids developed in \cite{LMSV-09}.

\item When $E=TQ$, equations (\ref{ec ge EL co}) are the standard $k$-cosymplectic geometric version of the Euler-Lagrange equations for field theories develop by M. de Le\'{o}n {\it et al} in \cite{LMS-2001}.

\item When $L$ does not depends on $\mathbf{t}$  and  $E=TQ$ and $\rho=id_{TQ}$, equations (\ref{ec ge EL co}) coincide with the Euler-Lagrange equations of the G\"{u}nther formalism \cite{Gu-1987}.

\end{enumerate}
\end{remark}

In the following table we write the geometric Lagrangian equations in the above particular cases.

\begin{center}
\begin{tabular}{|c|c|}
\hline
 & \begin{tabular}{c}
 {\sc Lagrangian formalism}\\
 Geometric Lagrangian equations
 \end{tabular}\\\hline
 \begin{tabular}{c}
 $k$-cosymplectic formalism\\ on Lie algebroids
 \end{tabular} & $\begin{array}{c}\\
 \mathcal{Y}^B(\xi_A)=\delta_A^B\\\noalign{\medskip}
    \displaystyle\sum_{A=1}^k\imath_{\xi_A}\Omega_L^A =\d^{\tec} E_L+
\displaystyle\sum_{A=1}^k\frac{\partial L}{\partial t^A}\mathcal{Y}_A\\\noalign{\medskip}
(\xi_1,\ldots, \xi_k) \makebox{\,family of } k \makebox{ sections of } \tec\\\quad
 \end{array}$\\\hline
  \begin{tabular}{c}
 $k$-symplectic  formalism\\ on Lie algebroids\\$\left(\frac{\partial L}{\partial t^A}=0,\, A=1,\ldots, k\right)$
 \end{tabular} & $\begin{array}{c}\\
     \displaystyle\sum_{A=1}^k\imath_{\xi_A}\Omega_L^A =\d^{\te} E_L\\\noalign{\medskip}
(\xi_1,\ldots, \xi_k) \makebox{\,family of } k \makebox{ sections of } \te\\\quad
 \end{array}$
 \\\hline
 \begin{tabular}{c}Standard\\
  $k$-cosymplectic  formalism\\ $(E=TQ)$\end{tabular} & $\begin{array}{c}\\
 dt^A(Y_B)=\delta^A_B \\  \noalign{\medskip}
 \displaystyle\sum_{A=1}^k \, i_{Y_A} \omega_L^A =\,
\d E_L + \,\displaystyle\sum_{A=1}^k\displaystyle\frac{\partial L}{\partial t^A}dt^A  \\  \noalign{\bigskip}
(Y_1,\dots,Y_k) \,\, \mbox{k-vector field on} \,\, \rk\times T^1_kQ\\\quad
 \end{array}$
  \\\hline
  \begin{tabular}{c}Standard\\
  $k$-symplectic  formalism\\ $(E=TQ)$\\$\left(\frac{\partial L}{\partial t^A}=0,\, A=1,\ldots, k\right)$\end{tabular} & $\begin{array}{c}\\
   \displaystyle\sum_{A=1}^k \, i_{Y_A} \omega_L^A =\,
\d E_L  \\  \noalign{\bigskip}
(Y_1,\dots,Y_k) \,\, \mbox{k-vector field on} \,\,   T^1_kQ\\\quad
 \end{array}$
  \\\hline
\end{tabular}
\end{center}

\begin{remark} When  $k=1$,
\begin{enumerate}
\item If $L$ explicitly depends on $t$,  equations (\ref{ec ge EL co}) are the equations of Lagrangian mechanics for time-dependent system defined on Lie algebroids, see for instance \cite{SMM-2002,SMM-2002(2)}. Moreover. In this case, when $E=TQ$, (\ref{ec ge EL co}) are the dynamical equations of  non-autonomous mechanics (see  \cite{EMR-1991}).

\item If $L$ does not depends on $\mathbf{t}$,  equations (\ref{ec ge EL co}) are the geometric equations for  autonomous lagrangian mechanics on Lie algebroids, see for instance \cite{Mart-2001(2)}. Finally in this case if $E=TQ$  we have the classical equations for autonomous mechanics.

\end{enumerate}
\end{remark}

\subsection{Hamiltonian formalism}\label{Ha}

In this section we extend the standard Hamiltonian $k$-cosymplectic formalism to Lie algebroids. In the following, we consider a Lie algebroid  $(E,\lcf\cdot,\cdot\rcf_E,\rho_E)$ over a manifold $Q$, and the dual bundle,
$\tau^{\;*}:E^{\;*}\to Q$  of $E$.

\subsubsection{The manifold $\rk\times\stackrel{k}{\oplus} E^*$.} The appropriate space   of the standard Hamiltonian $k$-cosymplectic formalism is the bundle  $\rktkqh$, where $\tkqh$ is the bundle of $k^1$-velocities of $Q$, that is, the Whitney sum of $k$ copies of $T^*Q$. For this generalization  to Lie algebroids,  it is natural to consider that the analog of $\rktkqh$ is $$\rk\times\keh\equiv\rk\times (E^*\oplus\stackrel{k}{\ldots}\oplus E^*)\,,$$ with the projection map
$$\widetilde{\mathrm{p}}^*\colon \rkeh\to Q,\qquad \widetilde{\mathrm{p}}^* (t^1,\ldots, t^k,{e_1}^*_{q},\ldots, {e_k}^*_{q})= {q}\,,$$
  $\keh$ being  the Whitney sum of $k$ copies of the dual space $E^*$.

Let us observe that the  elements of $\rkeh$ are of the form
$$(\mathbf{t} ,\mathbf{e}^*_{q})=(t^1,\ldots, t^k,{e_1}^*_{q},\ldots, {e_k}^*_{q})\,.$$

If $(q^i,y_\alpha)$ are local coordinates on
$(\tau^*)^{\,-1}(U)\subseteq E^*$, then the induced
 local coordinates $(t^A,q^i,y_\alpha^A)$ on $(\widetilde{\mathrm{p}}^*)^{\,-1}(U)\subseteq\rk\times\keh$ are given by
\begin{equation}\label{cke}
t^A(\mathbf{t} ,\mathbf{e}^*_{q})=t^A(\mathbf{t}),\quad
q^i(\mathbf{t} ,\mathbf{e}^*_{q})=q^i({q}),\quad
y_\alpha^A(\mathbf{t} ,\mathbf{e}^*_{q})=y_\alpha({e_A}^*_{q})\,. \end{equation}

\subsubsection{The Hamiltonian prolongation}\label{Ham prol}
We next consider the prolongation of a Lie algebroid $E$ over the fibration $\widetilde{\mathrm{p}}^*\colon \rk\times\keh\to Q$, that is (see section \ref{prolong})
\begin{equation}\label{tech}\tech=\{(a_{q},v_{(\mathbf{t},\mathbf{e}^*_q)})\in E\times T(\rkeh) /\;
\rho(a_{q})=T\widetilde{\mathrm{p}}^*(v_{(\mathbf{t},\mathbf{e}^*_q)})\}\,.\end{equation}

Taking into account the description of the prolongation $\mathcal{T}^EP$ and the results on Section
  \ref{prolong} (see also
\cite{CLMMM-2006,LMM-2005,Mart-2001}), we obtain
\begin{enumerate}
\item $\tech\subset  E\times T(\rkeh)$ is a Lie algebroid over $\rkeh$, with the projection $$
\widetilde{\tau}_{\rkeh}\colon  \tech\longrightarrow \rkeh$$ and Lie algebroid structure
$(\lcf\cdot,\cdot\rcf^{\widetilde{\mathrm{p}}^*},\rho^{\widetilde{\mathrm{p}}^*}\,)$,
where the anchor map
$$\rho^{\widetilde{\mathrm{p}}^*}\colon  \tech\to
T(\rkeh)$$ is the canonical projection onto the second factor. We refer to this Lie algebroid as the {\it $k$-cosymplectic Hamiltonian prolongation}

\item  Local coordinates $(t^A,q^i,y_\alpha^A)$ on $\rkeh$ induce local coordinates $(t^A,q^i,y_\alpha^A,z^\alpha,v_A,w_\alpha^A)$ on $\tech$, where
\begin{equation}\label{local coord cprol h}\begin{array}{lcllcl}
t^A(a_{q},v_{(\mathbf{t},\mathbf{e}^*_q)}) &=& t^A(\mathbf{t})
  \;,\quad & z^\alpha(a_{q},v_{(\mathbf{t},\mathbf{e}^*_q)})&=&y^\alpha(a_{q})
\;,\\\noalign{\medskip}
 q^i(a_{q},v_{(\mathbf{t},\mathbf{e}^*_q)}) &=& q^i({q})
  \;,\quad &
  v_A(a_{q},v_{(\mathbf{t},\mathbf{e}^*_q)})&=&v_{(\mathbf{t},\mathbf{e}_q^*)}(t^A)\;,\\\noalign{\medskip}
y_\alpha^A(a_{q},v_{(\mathbf{t},\mathbf{e}^*_q)})&=&y_\alpha^A(\mathbf{t},\mathbf{e}^*_{q})\;,
\quad & w^A_\alpha(a_{q},v_{(\mathbf{t},\mathbf{e}^*_q)})
&=&v_{(\mathbf{t},\mathbf{e}_q^*)}(y_\alpha^A)\;. \\
\end{array}
\end{equation}

\item  The set $\{\mathcal{Y}_A,\mathcal{X}_\alpha,\mathcal{V}_A^\alpha\}$ given by \[\mathcal{Y}_A,\,\mathcal{X}_\alpha,\,\mathcal{V}_A^\alpha\colon  \rkeh\to\mathcal{T}^E(\rkeh)\]
{\small\begin{equation}\label{basehc} \begin{array}{c}
\mathcal{Y}_A(\mathbf{t},\mathbf{e}_q^*)=
 (0_{q};\displaystyle\frac{\partial}{\partial
t^A}\Big\vert_{(\mathbf{t},\mathbf{e}_q^*)})\quad ,  \quad
\mathcal{X}_\alpha(\mathbf{t},\mathbf{e}_q^*) =
(e_\alpha({q});\rho^i_\alpha({q})\displaystyle\frac{\partial
}{\partial q^i}\Big\vert_{(\mathbf{t},\mathbf{e}_q^*)})\quad ,  \quad
%\\\noalign{\bigskip}
\mathcal{V}_A^\alpha(\mathbf{t},\mathbf{e}_q^*)
=(0_{q};\displaystyle\frac{\partial}{\partial
y_\alpha^A}\Big\vert_{(\mathbf{t},\mathbf{e}_q^*)})
\end{array}\end{equation}} is a local basis of  $\sec{\tech}$, the set of sections of  $\widetilde{\tau}_{\rkeh}$ (see (\ref{base k-prol})).

\item The anchor map $\rho^{\widetilde{\mathrm{p}}^*}\colon  \tech\to
T(\rkeh)$ allows us to associate a vector field with each section $\xi\colon \rkeh\to\tech$ of $\widetilde{\tau}_{\rkeh}$. Locally, if $\xi$ is given by
$$\xi=\xi^A\mathcal{Y}_A+\xi^\alpha\mathcal{X}_\alpha+\xi_\alpha^A\mathcal{V}_A^\alpha\in \sec{\tech},$$
then the associate vector field is
\begin{equation}\label{rholksimhc}
\rho^{\widetilde{\mathrm{p}}^*}(\xi)=\xi^A\derpar{}{t^A}+\rho^i_\alpha
\xi^\alpha\derpar{}{q^i} + \xi_\alpha^A\derpar{}{y_\alpha^A}\in
\vf(\rkeh)\,.\end{equation}

\item The Lie bracket of two sections of $\widetilde{\tau}_{\rkeh}$ is characterized by the relations (see (\ref{lie brack k-prol})),
\begin{equation}\label{lie brack tech}\hspace{-0.5cm}\begin{array}{lclclclclcl}
\lcf\mathcal{Y}_A,\mathcal{Y}_B\rcf^{\widetilde{\mathrm{p}}^*}&=&
0\;, &\quad &
\lcf\mathcal{Y}_A,\mathcal{X}_\alpha\rcf^{\widetilde{\mathrm{p}}^*}&=&
0\;, &\quad
&\lcf\mathcal{Y}_A,\mathcal{V}_B^\beta\rcf^{\widetilde{\mathrm{p}}^*}&=&
0\;,
\\\noalign{\medskip}
\lcf\mathcal{X}_\alpha,\mathcal{X}_\beta\rcf^{\widetilde{\mathrm{p}}^*}&=&
\mathcal{C}^\gamma_{\alpha\beta}\mathcal{X}_\gamma\;, &\quad &
\lcf\mathcal{X}_\alpha,\mathcal{V}_B^\beta\rcf^{\widetilde{\mathrm{p}}^*}&=&0\;,
&\quad &
\lcf\mathcal{V}_A^\alpha,\mathcal{V}_B^\beta\rcf^{\widetilde{\mathrm{p}}^*}&=&0\,.
\end{array}\end{equation}

\item If $\{\mathcal{Y}^A,\mathcal{X}^\alpha,
\mathcal{V}_\alpha^A\}$ is the dual basis of
$\{\mathcal{Y}_A,\mathcal{X}_\alpha, \mathcal{V}^\alpha_A\}$.  then the exterior differential is given by
\begin{equation}\label{ext dif tech}
\begin{array}{l}
\d^{\tech}f=\derpar{f}{t^A}\mathcal{Y}^A+\rho^i_\alpha\derpar{f}{q^i}\mathcal{X}^\alpha
+ \derpar{f}{y_\alpha^A}\mathcal{V}^A_\alpha\,,\quad \makebox{ for
all }  \; f \in \mathcal{C}^\infty(\rkeh)\\\noalign{\medskip}
\d^{\tech}\mathcal{Y}^A =0\quad , \quad
\d^{\tech}\mathcal{X}^\gamma =
-\displaystyle\frac{1}{2}\mathcal{C}^\gamma_{\alpha\beta}\mathcal{X}^\alpha\wedge\mathcal{X}^\beta
\quad , \quad
\d^{\tech}\mathcal{V}_\gamma^A =0\,,\end{array}\end{equation} (see (\ref{dtp})).
\end{enumerate}

\begin{remark}\label{remark equiv h}
In the particular case $E= TQ$, the manifold $\mathcal{T}^E(\rkeh)$ reduces to $T(\rktkqh)$. The proof is analogous to the on in  remark \ref{remark equivch}. \end{remark}

%%%%
\subsubsection{The vector bundle
$\tech\oplus\stackrel{k}{\ldots}\oplus \tech$.}

In the standard Hamiltonian $k$-cosymplectic formalism one obtains the solutions of the Hamilton equations as integral sections of certain $k$-vector fields on $\rktkqh$, that is sections of
$$\tau^k_{\rktkqh}\colon T^1_k(\rk\times (T^1_k)^*Q) \to \rk\times (T^1_k)^*Q\,.$$

Since on Lie algebroids the vector bundle $\mathcal{T}^E(\rkeh)$ plays the role of $T(\rktkqh)$, it is natural to assume that the role of
$$T^1_k(\rktkqh)\equiv T(\rktkqh)\oplus\stackrel{k}{\ldots}\oplus
T(\rktkqh)$$ is played by
$$\tkech\colon =\tech\oplus\stackrel{k}{\ldots}\oplus \tech\,,$$
the Whitney sum of k copies of $\mathcal{T}^E(\rkeh)$, being  the canonical projection $\widetilde{\tau}^{\,k}_{\rkeh}\colon \tkech\to \rkeh\,$ given by
$$ \widetilde{\tau}^{\,k}_{\rkeh}(Z^1_{ (\mathbf{t},{\mathbf{e}}_{q}^*)},\ldots, Z^k_{ (\mathbf{t},{\mathbf{e}}_{q}^*)})=(\mathbf{t},{\mathbf{e}}_{q}^*),$$where
$Z^A_{ (\mathbf{t},{\mathbf{e}}_{q}^*)}=({a_A}_q,{v_A}_{ (\mathbf{t},{\mathbf{e}}_{q}^*)})\in \mathcal{T}^E(\rkeh),\,A=1,\ldots, k.$
We have the following

%We now prove that there exists a $k$-vector field on $\rkeh$ associated with each section $\xi$ of $\tau^k_{\rktkqh}$. Note that to give a section
%$$\mathbf{\xi}\colon \rkeh\to (\mathcal{T}^E)^1_k(\rkeh))=\tech\oplus\stackrel{k}{\ldots}\oplus \tech$$ of $\tau^k_{\rktkqh}$ is equivalent to giving $k$ sections $\xi_1,\ldots, \xi_k$ of the Hamiltonian prolongation $\mathcal{T}^E(\rkeh)$, namely the projections of $\xi$ on each summand $\mathcal{T}^E(\rkeh)$.

\begin{proposition}
Let $\xi=(\xi_1,\ldots,\xi_k)$ be a section of $\tau^k_{\rktkqh}$. Then
$$(\rho^{\tilde{p}^{\,*}}(\xi_1),\ldots,\rho^{\tilde{p}^{\,*}}(\xi_k))\colon \rkeh\to
T^1_k(\rkeh)$$
$$\xi_A\colon  \rkeh\to \tech$$ is a $k$-vector field on $\rkeh$, where $\rho^{\tilde{p}^{\,*}}$ is the anchor map of the Lie algebroid $\mathcal{T}^E(\rkeh)$.
\end{proposition}
\proof Directly from (\ref{rholksimhc}) and the above remark.\qed

\subsubsection{Hamiltonian formalism} Let $(E,\lcf\cdot,\cdot\rcf_E,\rho_E)$ be a Lie algebroid on a manifold $Q$,
and $H:\rk\times\stackrel{k}{\oplus} E^{\;*}\to \r$  a Hamiltonian function. To develop the Hamiltonian $k$-cosymplectic formalism on Lie algebroids, we need to define an appropriate notion of Liouville sections.
\paragraph{\bf The Liouville sections} The {\it Liouville $1$-sections} are defined as  sections of the bundle
$\big(\mathcal{T}^E(\rkeh)\big)^{\,*}\to\rkeh$ such that
$$\begin{array}{rcc}
\Theta^A:\rkeh& \longrightarrow & (\mathcal{T}^E(\rkeh))^{\;*}
\\\noalign{\medskip}
(\mathbf{t},\mathbf{e}_{q}^{\;*} )& \longmapsto & \Theta^A_{(\mathbf{t},\mathbf{e}_{q}^{\;*} )}\end{array}\quad 1\leq A\leq k\,,
 $$
where $\Theta^A_{(\mathbf{t},\mathbf{e}_{q}^{\;*} )}\colon (\mathcal{T}^E(\rkeh) )_{(\mathbf{t},\mathbf{e}_{q}^{\;*} )} \to \r$ is the linear function:
 \begin{equation}\label{theta A kc}{\begin{array}{rll}
 ( a_{q}, v_{(\mathbf{t},\mathbf{e}_{q}^{\;*} )}) & \longmapsto &
  \Theta^A_{(\mathbf{t},\mathbf{e}_{q}^{\;*} )}( a_{q}, v_{(\mathbf{t},\mathbf{e}_{q}^{\;*} )})=e_{A_{q}}^{\;*}(a_{q})\;,
\end{array}}\end{equation}
for each $ a_{q}\in E,\,(\mathbf{t},\mathbf{e}_{q}^{\;*} )=(\mathbf{t},{e_1}_{q}^*,\ldots,{e_k}_{q}^*)\in \rkeh$ and $ v_{(\mathbf{t},\mathbf{e}_{q}^{\;*} )}\in T_{(\mathbf{t},\mathbf{e}_{q}^{\;*} )}(\rkeh)$. The {\it Liouville $2$-sections}
$$\Omega^A:\rkeh \to
\Lambda^2\big[\mathcal{T}^E(\rkeh)\big]^{\;*}\;,\quad1\leq A\leq k$$  defined by
$$
\Omega^A=- \d^{\tech}\Theta^A\;,
$$ where $\d^{\tech}$ denotes the exterior differential  on the Lie algebroid  $\tech$ ( see  (\ref{ext dif tech})).

Locally, if
$\{\mathcal{Y}_B,\mathcal{X}_\alpha,\;\mathcal{V}_B^\beta\}$ is a local basis of $\Sec{\mathcal{T}^E(\rkeh)}$ and
$\{\mathcal{Y}^B,\mathcal{X}^\alpha,\;\mathcal{V}^B_\beta\}$ its dual basis, then from (\ref{basehc}),
\begin{equation}\label{theta*kcsim}
\Theta^A=\displaystyle\sum_{\beta=1}^my^A_\beta\mathcal{X}^\beta\;,\quad 1\leq
A\leq k\,,
\end{equation} and from (\ref{lie brack tech}), (\ref{ext dif tech}) and (\ref{theta*kcsim}),
 \begin{equation}\label{omega A*kcsim}
\Omega^A=\sum_{\beta}\mathcal{X}^\beta\wedge\mathcal{V}^A_\beta +
\displaystyle\frac{1}{2} \sum_{\beta,\gamma,\delta}
y^A_\delta\mathcal{C}^\delta_{\beta\gamma}\mathcal{X}^\beta\wedge
\mathcal{X}^\gamma\;, \quad 1\leq A\leq k\;.
\end{equation}
\begin{remark} When
  $E= TQ$ and $\rho=id_{TQ}$
then $$\Omega^A(X,Y)=\omega^A(X,Y)\,,\qquad 1\leq A\leq k\,,$$
where $X,Y$ are vector field on $\rktkqh$ and $\omega^1,\ldots,$
$ \omega^k$ are the canonical $2$-forms of the standard Hamiltonian $k$-cosymplectic formalism (see (\ref{locexp})).
\end{remark}

\paragraph{\bf The Hamiltonian equations.}\label{Sec 8.2.2.}

\begin{theorem}\label{alhamform co}
Let $H:\rk\times\stackrel{k}{\oplus}E^*\to \r$ be a Hamiltonian function and
$$\xi_H=(\xi_1,\ldots,\xi_k):\rk\times\stackrel{k}{\oplus}E \to
\tkech\cong \tech\oplus\stackrel{k}{\ldots}\oplus \tech$$ a section of
$\widetilde{\tau}^{\,k}_{\rk\times\stackrel{k}{\oplus}E^*}$, (or equivalently, $\xi_1,\ldots,\xi_k$ are $k$ sections of the Hamiltonian prolongation), such that
\begin{equation}\label{geometric ec H al}
 \mathcal{Y}^B(\xi_A)=\delta_A^B\quad,\quad
    \displaystyle\sum_{A=1}^k\imath_{\xi_A}\Omega^A =\d^{\tech} H-
\displaystyle\sum_{A=1}^k\frac{\partial H}{\partial
t^A}\mathcal{Y}^A\;.\end{equation}

 If
$\psi:\rk\to\rk\times \stackrel{k}{\oplus}E^*,\,\psi(\mathbf{t}) =
(\mathbf{t},\psi^i(\mathbf{t}),\psi^A_\alpha(\mathbf{t}))$  is an integral section of $\xi_H$, then $\psi$ is a solution of the following system of partial differential equations:
\begin{equation} \label{Hamilton eq}
\begin{array}{rcl}\derpar{\psi^{\;i}}{t^A}\Big\vert_{\mathbf{t}}
&=&\rho^i_\alpha\displaystyle\frac{\partial H}{\partial
y^A_\alpha}\Big\vert_{\psi(\mathbf{t})}\;,\\\noalign{\medskip}
\displaystyle\sum_{A=1}^k\derpar{\psi^{\;A}_\beta}{t^A}\Big\vert_{\mathbf{t}} &=& -
\Big(
 \rho^i_\beta\displaystyle\frac{\partial H}{\partial
q^i}\Big\vert_{\psi(\mathbf{t})}+\displaystyle\sum_{A=1}^k
\psi^A_\gamma(\mathbf{t})\mathcal{C}^\gamma_{\alpha\beta}
\displaystyle\frac{\partial H}{\partial
y^A_\alpha}\Big\vert_{\psi(\mathbf{t})}\Big)\,.
\end{array}\end{equation}
\end{theorem}

\begin{remark}
{\rm In the particular case $E=TQ$ and $\rho=id_{TQ}$,  equations (\ref{Hamilton eq}) are the Hamilton field equations. Accordingly, equations (\ref{Hamilton eq}) are called {\it the Hamilton equations for Lie algebroids}.}
\end{remark}

\proof
The proof is analogous to that of Theorem \ref{algeform} in  section
 \ref{Sec 8.1.3.}. A schedule of this proof is the following:

Consider $\{\mathcal{Y}_B,\mathcal{X}_\alpha,\;\mathcal{V}_B^\beta\}$, a local basis of sections of $
\widetilde{\tau}_{\rkeh}\colon  \tech\longrightarrow \rkeh$. If $\xi_H=(\xi_1,\ldots,\xi_k)$, then each component  $\xi_A$  can be written in the form
\begin{equation}\label{localxia}\xi_A=\xi_A^B\mathcal{Y}_B+\xi_A^\alpha\mathcal{X}_\alpha+
(\xi_A)_\alpha^B\mathcal{V}^\alpha_B\,.\end{equation} and from  (\ref{ext dif tech}), (\ref{omega A*kcsim}) and (\ref{localxia}) the local expression of
   (\ref{geometric ec H al}) is
\begin{equation}\label{localxi-3}
\xi_A^B = \delta_{A}^B\quad,\quad
 \xi^\alpha_A= \displaystyle\frac{\partial H}{\partial y^A_\alpha}\quad
,\quad \displaystyle\sum_{A=1}^k(\xi_A)^A_\alpha= - \Big(
\rho^i_\alpha\displaystyle\frac{\partial H}{\partial q^i}+\displaystyle\sum_{A=1}^k
\mathcal{C}^\gamma_{\alpha\beta}y^A_\gamma \displaystyle\frac{\partial
H}{\partial y^A_\beta}\Big) \,.
\end{equation}

Also, if $\psi\colon\rk\to\rkeh, \;\psi(\mathbf{t}) =
(\mathbf{t},\psi^i(\mathbf{t}),\psi^A_\alpha(\mathbf{t}))$ is an integral section of $\xi_H$, that is $\psi$ is an integral section of $(\rho^{\tilde{p}^{\,*}}(\xi_1),\ldots,\rho^{\tilde{p}^{\,*}}(\xi_k))$, the associated $k$-vector field on $\rkeh$, then
\begin{equation}\label{sint1}
 \rho^i_\alpha(\xi^\alpha_A\circ\psi) = \displaystyle\frac{\partial \psi^i}{\partial
 t^A}\;,\;
  (\xi_A)_\beta^B\circ\psi = \displaystyle\frac{\partial \psi^B_\beta}{\partial
  t^A}\;.
\end{equation}

>From (\ref{localxi-3}) and (\ref{sint1}), $$
  \displaystyle\frac{\partial \psi^i}{\partial t^A} = \displaystyle\frac{\partial H}
  {\partial y^A_\alpha}\rho^i_\alpha \quad\makebox{and}\quad
  \displaystyle\sum_{A=1}^k  \displaystyle\frac{\partial \psi^A_\alpha}{\partial
  t^A} =-\left(\mathcal{C}^\delta_{\alpha\beta}\,\psi^A_\delta\,
  \derpar{H}{y^A_\beta}
+ \rho^i_\alpha \displaystyle\frac{\partial H}{\partial q^i} \right) \,.
$$
\qed
\begin{remark}\
\begin{enumerate}
\item When $H$ does not depends on $\mathbf{t}$, then it can be considered as a map $H\colon \keh\to \r$. In this case, the sections $\Omega^A$ can be thought as sections of $\mathcal{T}^E(\keh)$ and from (\ref{geometric ec H al}) one obtains the $k$-symplectic Hamiltonian equations for field theories on Lie algebroids (see \cite{LMSV-09}).

\item When $E=TQ$ and $\rho=id_{TQ}$, equations  (\ref{geometric ec H al}) are the standard $k$-cosymplectic geometric version of the Hamilton equations for field theories develop by M. de Le\'{o}n {\it et al} in \cite{LMORS-1998}.

\item If $H$ does not depends on $\mathbf{t}$  and we consider the case  $E=TQ$ and $\rho=Id_{TQ}$ we obtain the standard $k$-symplectic geometric version of the Hamilton equations for field theories (see, for instance \cite{MRS-2004,{RSV-2007}}).

\end{enumerate}
\end{remark}

In the following table we write the geometric Lagrangian equations in the above particular cases.
\begin{center}
\begin{tabular}{|c|c|}
\hline
 & \begin{tabular}{c}
 {\sc Hamiltonian formalism}\\
 Geometric Hamiltonian equations
 \end{tabular}\\\hline
 \begin{tabular}{c}
 $k$-cosymplectic formalism\\ on Lie algebroids
 \end{tabular} & $\begin{array}{c}\\
 \mathcal{Y}^B(\xi_A)=\delta_A^B\\\noalign{\medskip}
    \displaystyle\sum_{A=1}^k\imath_{\xi_A}\Omega^A =\d^{\tech} H-
\displaystyle\sum_{A=1}^k\frac{\partial H}{\partial
t^A}\mathcal{Y}^A\\\noalign{\medskip}
(\xi_1,\ldots, \xi_k) \makebox{\,family of } k \makebox{ sections of } \tech\\\quad
 \end{array}$\\\hline
  \begin{tabular}{c}
 $k$-symplectic  formalism\\ on Lie algebroids \\$\left(\frac{\partial H}{\partial t^A}=0\,,A=1,\ldots, k\right)$
 \end{tabular} & $\begin{array}{c}\\
     \displaystyle\sum_{A=1}^k\imath_{\xi_A}\Omega^A =\d^{\teh} H\\\noalign{\medskip}
(\xi_1,\ldots, \xi_k) \makebox{\,family of } k \makebox{ sections of } \teh\\\quad
 \end{array}$
 \\\hline
 \begin{tabular}{c}Standard\\
  $k$-cosymplectic  formalism\\ $(E=TQ)$\end{tabular} & $\begin{array}{c}\\
 dt^A(Y_B)=\delta^A_B \\  \noalign{\medskip}
 \displaystyle\sum_{A=1}^k \, i_{Y_A} \omega^A =\,
\d H - \,\displaystyle\sum_{A=1}^k\displaystyle\frac{\partial H}{\partial t^A}dt^A  \\  \noalign{\bigskip}
(Y_1,\dots,Y_k) \,\, \mbox{k-vector field on} \,\, \rk\times (T^1_k)^*Q\\\quad
 \end{array}$
  \\\hline
  \begin{tabular}{c}Standard\\
  $k$-symplectic  formalism\\ $(E=TQ)$\\$\left(\frac{\partial H}{\partial t^A}=0\,,A=1,\ldots, k\right)$\end{tabular} & $\begin{array}{c}\\
   \displaystyle\sum_{A=1}^k \, i_{Y_A} \omega_L^A =\,
\d H  \\  \noalign{\bigskip}
(Y_1,\dots,Y_k) \,\, \mbox{k-vector field on} \,\,   (T^1_k)^*Q\\\quad
 \end{array}$
  \\\hline
\end{tabular}
\end{center}

\begin{remark} When $k=1$,
\begin{enumerate}
\item If $H$ explicitly depends on $t$,   equations (\ref{geometric ec H al}) are the equations of Hamiltonian mechanics for time-dependent system defined on Lie algebroids ( see \cite{SMM-2002,SMM-2002(2)}, for instance). Moreover, when $E=TQ$ and $\,\rho=Id_{TQ}$  we have the dynamical equations of the non-autonomous mechanics (see  \cite{EMR-1991}).

\item If $H$ does not depends on $t$ ,  equations (\ref{geometric ec H al}) are the geometric equations of autonomous Hamiltonian mechanics on Lie algebroids (see, for instance, \cite{Mart-2001(2)}). In this case,  if $E=TQ$ and $\rho=id_{TQ}$ we have the classical equations of autonomous mechanics.

\end{enumerate}
\end{remark}

\subsection{Equivalence between the Lagrangian and Hamiltonian formalism}
  \label{eq}

In the standard case the Hamiltonian and Lagrangian $k$-cosymplectic formulations are equivalents when the Lagrangian is hyperregular. On the $k$-symplectic formalism on Lie algebroid we have obtained a similar result (see \cite{LMSV-09}). In this section we will define the Legendre transformation on Lie algebroids and we will establish the equivalence between the Lagrangian and Hamiltonian formalisms when the Lagrangian function is hyperregular.

\begin{definition}
The {\rm Legendre transformation} associated with $L:\rke\to\r$ is the smooth map
$$\Lec:\rke\to\rkeh$$
defined by
\[\Lec(\mathbf{t},\mathbf{e}_q)=\Big(\mathbf{t},[\Lec(\mathbf{t},\mathbf{e}_q)]^1,\ldots, [\Lec(\mathbf{t},\mathbf{e}_q)]^k\Big)\] where
\[[\Lec(\mathbf{t},\mathbf{e}_q)]^A(u_{q})=\displaystyle\frac{d}{ds}\Big\vert_{s=0}L(\mathbf{t},e_{1_{q}},\ldots,
{e_A}_{q}+su_{q},\ldots,
e_{k_{q}})\,, \quad 1\leq A\leq k\; ,\] where $u_{q}\in
E_{q}$ and $(\mathbf{t},\mathbf{e}_q)=(\mathbf{t},e_{1_{q}},\ldots, e_{k_{q}})\in \rke$.
\end{definition}

The map $\Lec$ is well defined, and its local expression is
\[\Lec(t^A,q^i,y^\alpha_A)=(t^A,q^i,\displaystyle\frac{\partial L}{\partial
y^\alpha_A})\,.\] From this expression, it is easy to prove that the Lagrangian $L$ is regular if and only if $\Lec$ is a local diffeomorphism.

\begin{remark} When
$E=TQ$, the Legendre transformation defined here coincides with the Legendre map of the standard $k$-cosymplectic formalism, see \cite{LMS-2001,{MSV-2005}}. \end{remark}

$\Lec$ induces a map
$$\mathcal{T}^E\Lec:\tec\to
\tech$$ defined by
\[\mathcal{T}^E\Lec(a_{q},\mathrm{v}_{(\mathbf{t},\mathbf{e}_{q})})=
\Big(a_{q},(\Lec)_*(\mathbf{t},\mathbf{e}_{q})(\mathrm{v}_{(\mathbf{t},\mathbf{e}_{q})})
\Big)\,,\]  where $a_{q}\in
E_{q},\;(\mathbf{t},\mathbf{e}_{q})\in\rk\times
\stackrel{k}{\oplus}E$ and
$(a_{q},\mathrm{v}_{(\mathbf{t},\mathbf{e}_{q})})\in
\tec\subset  E\times T(\rke).$

\begin{theorem}\label{equivforma al}The pair $(\mathcal{T}^E\Lec, \Lec)$ is a morphism between the Lie algebroid
$(\mathcal{T}^E(\rk\times\stackrel{k}{\oplus}E),\rho^{\widetilde{\mathrm{p}} },\lcf\cdot,\cdot\rcf^{\widetilde{\mathrm{p}} })$
and
$(\mathcal{T}^E(\rk\times\stackrel{k}{\oplus}E^*),\rho^{\widetilde{\mathrm{p}} ^{\,*}},\lcf\cdot,\cdot\rcf^{\widetilde{\mathrm{p}} ^{\,*}})$.

\[\xymatrix@R=15mm{\mathcal{T}^E(\rk\times\stackrel{k}{\oplus}E)
\ar[rr]^-{\mathcal{T}^E\Lec}\ar[d]_-{\widetilde{\tau}_{\rk\times\stackrel{k}{\oplus}E}}
&&\mathcal{T}^E(\rk\times\stackrel{k}{\oplus}E^{\,*})\ar[d]^-{\widetilde{\tau}_{\rkeh}}\\
\rk\times\stackrel{k}{\oplus}E\ar[rr]_-{\Lec}
&&\rk\times\stackrel{k}{\oplus}E^{\,*} }\]

Moreover, if  $\Theta_L^A$ and
$\Omega_L^A$ are, respectively, the Poincar\'{e}-Cartan $1$-sections and $2$-sections associated with $L\colon\rke\to\r$, and $\Theta^A$ and $\Omega^A$, respectively, the  Liouville $1$-sections and $2$-sections on
$\mathcal{T}^E(\rk\times\stackrel{k}{\oplus}E^*)$), then
\begin{equation}\label{equiv formas}
(\mathcal{T}^E\Lec, \Lec)^*\Theta^A=\Theta_L^A,\qquad
(\mathcal{T}^E\Lec, \Lec)^*\Omega^A=\Omega_L^A\,,\quad 1\leq A\leq
k\,.
\end{equation}
\end{theorem}

\proof The proof is analogous to the one in the   $k$-symplectic case, see Theorem 4.30 in \cite{LMSV-09}.\qed

\begin{remark}
When
$E=TQ$ and $\rho=id_{TQ}$,  it establishes the relation between the Lagrangian and Hamiltonian formalism in the standard $k$-cosymplectic approach (see \cite{LMS-2001}). \end{remark}

We next assume that $L$ is {\it hyperregular}, that is, that $\Lec$ is a global diffeomorphism. In this case we may consider the Hamiltonian function $H\colon \rkeh\to \r$ defined by  $$H=E_L\circ (\Lec)^{\,-1},$$ where $E_L$ is the energy function associated with $L$, given by (\ref{local enerco}), and
$(\Lec)^{\,-1}$ is the inverse of the Legendre transformation.
\[\xymatrix@=12mm{\rkeh\ar[r]^-{\Lec^{\,-1}}\ar@{-->}[rd]_-{H}
&\rke\ar[d]^-{E_L}\\ & \r}\]
%\begin{lemma}
%If the Lagrangian $L$ is hyperregular, then $\mathcal{T}^E\Lec$
%is a diffeomorphism.
%\end{lemma}

By a similar computation that in the Theorem 4.33 in \cite{LMSV-09}
we prove the following theorem, which establishes the equivalence between the Lagrangian and Hamiltonian $k$-cosymplectic formulations on Lie algebroids.
\begin{theorem} Let $L$ be a hyperregular Lagrangian. There is a bijective correspondence between the set of maps
 $\eta:\rk\to \rk\times\stackrel{k}{\oplus}E$ such that
$\eta$ is an integral section of a solution $\xi_L$ of the geometric Euler-Lagrange equations (\ref{ec ge EL co}) and the set of maps $\psi:\rk\to\rk\times \stackrel{k}{\oplus}E^{\,*}$
which are integral sections of some solution $\xi_H$ of the geometric Hamilton equations (\ref{Hamilton eq}).
\end{theorem}

\proof It is similar to the proof of the $k$-symplectic formalism on Lie algebroids, see Theorem  4.33 in \cite{LMSV-09}. Here we must only to take into account the relationship
between
$\xi_L=(\xi_L^1,\ldots,\xi_L^k)$ and $\xi_H=(\xi_H^1,\ldots,\xi_H^k)$
given by:
$$\xi_H^A\circ \Lec = \mathcal{T}^E\Lec\circ\;\xi_L^A\,,\quad
A=1,\ldots,k\,.$$\qed

\begin{remark}{\rm\
\begin{enumerate}
\item When $E=TQ$, this theorem establishes the equivalence between the $k$-cosymplectic Lagrangian and the Hamiltonian formalism (see \cite{LMS-2001,{MSV-2005}}).

\item When $L$ and $H$ do not depend on $\mathbf{t}$, the above Theorem reduces to the Theorem 4.33 in \cite{LMSV-09}, which establishes the equivalence between the Lagrangian and Hamiltonian formalism on Lie algebroids on the $k$-symplectic approach.
\end{enumerate}
}\end{remark}

\section{Examples}\label{examples}

\paragraph{\bf Harmonic maps, \cite{{CGR-2001},EL-78}}

Let us remember that a smooth map $\varphi\colon M\to N$ between Riemannian manifolds $(M,g)$ and $(N,h)$ is called \textit{harmonic} if it is a critical point of the energy functional $E$, which, when $M$ is  compact oriented manifold, is defined as $$E(\varphi)=\int_{M}\frac{1}{2}trace_g\varphi^*h\,dv_g$$
where $dv_g$ denotes the measure on $M$ induced by its metric and, in local coordinates, the expression $\frac{1}{2}trace_g\varphi^*h$ reads
$$\frac{1}{2} g^{ij}h_{\alpha\beta}\derpar{\varphi^\alpha}{x^i}\derpar{\varphi^\beta}{x^j}\; .$$

 This definition is extended to the case when $M$ is not compact requiring that the restriction of $\varphi$ to every compact domain to be harmonic.

Now we will consider the particular case $M=\rk$ and $N=G$ a Riemannian matrix Lie group.  In this case we denote the trivial principal fiber bundle by  $\pi\colon \rk\times G\to \rk$, and we identify ${\rm Sec} (\rk\times G)$ with $\mathcal{C}^\infty(\rk,G)$. For each $\phi\in\mathcal{C}^\infty(\rk,G)$, the Riemannian metrics on $\rk$ and $G$ naturally induce a metric $<\cdot,\cdot>$ on $\mathcal{C}^\infty (T^*\rk\otimes\varphi^*(TG))$, and so we may define the energy $E$ on $\mathcal{C}^\infty(\rk, M)$ by
\begin{equation}\label{energy-harmonic}E(\varphi)=\int_{\rk}L(\varphi^{[1]}(t))d^kt\end{equation}
where $L(\varphi^{[1]}(t))=\frac{1}{2}<T\varphi,T\varphi>$ and $d^kt=dt^1\wedge\ldots dt^k$ is the volume element of $\rk$.

The Euler-Lagrange equations for (\ref{energy-harmonic}) are given by (see, for example \cite{EL-78}),
\begin{equation}\label{EL eq Harmonic}
Trace(\nabla T\phi)=0\,,
\end{equation} where $\nabla$ is the induced Riemannian covariant derivative on $\mathcal{C}^\infty(T^*\rk\otimes \varphi^*(TG))$ and $Trace$ is the trace defined by $g$ (see, for example \cite{EL-78}). By definition, the set of harmonic maps from $\rk$ to $G$ is the subset of ${\rm Sec}(\rk\times G)$ whose elements solve (\ref{EL eq Harmonic}).

Using Einstein's summation convention, we have the following coordinate expressions:
\begin{equation}\label{Local Lagran harmonic}
\begin{array}{rccl}
L\colon & J^1(\rk\times G)\equiv\rk\times T^1_kG & \to & \r\\\noalign{\medskip}
 & (t^A,q^i,v^i_A) &\mapsto & L(t^A,q^i,v^i_A)=\frac{1}{2}g^{AB}v^i_Av^j_Bh_{ij}
 \end{array}
 \end{equation}
 where $t^A$ denoted the local coordinates on $\rk$, that is, the space-time coordinates, $q^i=\varphi^i$ the components of the field $\varphi$ and $v^i_A=\derpar{\varphi^i}{t^A}$ the partial derivatives of the components of the field. From (\ref{EL eq Harmonic}) one obtains
 \begin{equation}\label{Local EL eq Harmonic}
 g^{AB}\left(\derpars{\varphi^i}{t^A}{t^B}-\Gamma^C_{AB}\derpar{\varphi^i}{t^C}+\widetilde{\Gamma}^i
 _{jk}\derpar{\varphi^j}{t^A}\derpar{\varphi^k}{t^B}\right)=0\,\qquad 1\leq i\leq n\,,
 \end{equation}where $\Gamma^C_{AB}$ and $\widetilde{\Gamma}^i_{jk}$ denote the Christoffel symbols of the Levi-Civita connections of $g$ and $h$.

 We shall derive the reduced form of (\ref{EL eq Harmonic}) for two specific cases: $G=\r$ and $G= \mathbb{S}^3\cong SU(2)$. In general, one obtains,
\begin{equation}\label{connection bundle}\mathcal{C}(\rk\times G)\cong(J^1(\rk\times G)/G\cong(\rk\times T^1_kG)/G\cong(\rk\times G\times \mathfrak{g}\times\stackrel{k}{\ldots}\times \mathfrak{g})/G\cong\rk\times \mathfrak{g}\times\stackrel{k}{\ldots}\times \mathfrak{g}\,,\end{equation} where $\mathcal{C}(\rk\times G)\to \rk$ is the bundle of connections (see \cite{CRS-2000}).

For the case that $G=\r$, the abelian group of translations, from (\ref{connection bundle}) we obtain that $\mathcal{C}(\rk\times G)\cong\rk\times \rk$ and therefore, a section $\sigma$ of the bundle connections can be thought as a $1$-form on $\rk$ with local expression $\sigma=p_Adt^A$, where $(t^A,p_B)$  are local coordinates on $\rk\times \rk$.

The Lagrangian $L$ is clearly $\r$-invariant. Denoting by $\ell$ the projection of $L$ to $\mathcal{C}(P)\cong\mathcal{C}(\rk\times G)\cong\rk\times \rk$, in local coordinates, we obtain $\ell(t^A,p_b)= \frac{1}{2}g^{AB}p_Ap_B$. Now, we can write the Euler-Lagrange equations (\ref{eq E-L cosim}) for this Lagrangian $\ell$ and we obtain
\begin{equation}\label{E-P eq}\begin{array}{l}
\derpar{(g^{AB}p_B)}{t^A}+\Gamma^A_{AC}g^{CB}p_B=0\\\noalign{\medskip}
\derpar{p_A}{t^B}=\derpar{p_B}{t^A}\,.
\end{array}
\end{equation}
Let us observe that the first equation is the Euler-Poincar\'{e} equation for $\ell$ and the second equation is the condition of the vanishing curvature on the trivial connection for $\rk\times G$ (see, for instance,  \cite{CRS-2000}).

We denote by $q\colon \rk\times T^1_k\r\to(\rk\times T^1_k\r)/\r$ the canonical projection, let $\sigma=q(T\varphi)$, then $p_A=\partial \varphi /\partial t^A$, this condition together the equations (\ref{E-P eq}) is equivalent to (\ref{Local EL eq Harmonic}).

For the case $G=\mathbb{S}^3\cong SU(2)$, from (\ref{connection bundle}) we know that $\mathcal{C}(\rk\times SU(2))\cong\rk\times \mathfrak{su}(2)\times\stackrel{k}{\ldots}\times \mathfrak{su}(2)$ and we can  make the identification $$T^*\rk\otimes \mathfrak{su}(2)\cong\rk\times \mathfrak{su}(2)\times\stackrel{k}{\ldots}\times \mathfrak{su}(2).$$ This identification is  locally given as follow: Let $\{E_1,E_2,E_3\}$ be a basis of $\mathfrak{su}(2)$, then a section of $T^*\rk\otimes \mathfrak{su}(2)\to\rk$ can be written as $\sigma(t)=p^A_idt^A\otimes E_i$. This element $\sigma$ identifies with the element of $\rk\times \mathfrak{su}(2)\times\stackrel{k}{\ldots}\times \mathfrak{su}(2)$ with local coordinates $(t^A, p^A_i)$.

The lagrangian $L$, (see (\ref{Local Lagran harmonic})), is $\mathfrak{su}(2)$-invariant and its projection to $\rk\times \mathfrak{su}(2)\times\stackrel{k}{\ldots}\times \mathfrak{su}(2)$ is
$$\ell(t^A,p^A_i)=\frac{1}{2}g^{AB}p^A_ip^B_jh_{ij}\,.$$
Then the Euler-Lagrange equations (\ref{eq E-L cosim}) write, in this case, as follow:
\begin{equation}\label{harmonic EL su2}
\begin{array}{l}
\derpar{(g^{AB}p^B_ih_{ij})}{t^A} + \Gamma^C_{CB}g^{AB}p^A_ih_{ij}+g^{AB}p^A_ip^B_kc^l_{kj}h_{il}=0\\\noalign{\medskip}
\derpar{p^A_k}{t^B}-\derpar{p^B_k}{t^A} + p^A_ip^B_jc^k_{ij}=0\,.
\end{array}
\end{equation}

The first group of equations of (\ref{harmonic EL su2}) are the Euler-Poincar\'{e} equations for the trivial connection of $\rk\times SU(2)$, the second group of equations represents the  vanishing curvature condition  (see \cite{CRS-2000} for more details).

\paragraph{\bf Classical Euler-Poincar\'{e} equations} For a Lie Group $G$, we consider the principal fiber bundle $\pi\colon\r\times G\to\r$. Let $L\colon J^1(\r\times G)\cong\r\times TG\to \r$ be a $G$-invariant Lagrangian. Taking into account (\ref{connection bundle}) with $k=1$ we obtain the following identifications
$$\mathcal{C}(\r\times G)\cong (\r\times TG)/G\cong \r\times \mathfrak{g}\,.$$

In a similar way that in the above example we obtain if $\ell$ is the projection of $L$ to $\mathcal{C}(\r\times G)$, then the Euler-Lagrange equations associated to $\ell$ are the Classical Euler-Poincar\'{e} equations, see for instance \cite{CRS-2000} or \cite{MR-1994}.

\paragraph{\bf Systems with symmetry.}

Consider a principal bundle $\pi: \bar{Q}\longrightarrow {Q}=\bar{Q}/G$. Let $A: T\bar{Q}\longrightarrow {\mathfrak g}$ be a fixed  principal connection with curvature  $B: T\bar{Q}\oplus T\bar{Q}\longrightarrow {\mathfrak g}$.
The connection $A$ determines an isomorphism
between the vector bundles $T\bar{Q}/G\to Q$ and  $TQ\oplus \widetilde{\mathfrak g}\longrightarrow Q$, where $\widetilde{\mathfrak g}=(\bar{Q}\times {\mathfrak g})/G$ is the adjoint bundle (see \cite{CMR-2001}):
\[
[ {v}_{\bar{  q}}] \leftrightarrow T_{\bar{{  q}}}\pi( {v}_{\bar{{  q}}})\oplus [(\bar{{  q}}, A( {v}_{\bar{{  q}}}))]
\]
 where $ {v}_{\bar{{  q}}}\in T_{\bar{{  q}}}\bar{Q}$. The connection allows us to obtain a local basis of sections of ${\rm Sec} (T\bar{Q}/G)={\mathfrak X}(Q)\oplus {\rm Sec}( \widetilde{{\mathfrak g}})$ as follows. Let ${\mathfrak e}$ be the identity element of the Lie group $G$ and assume that there are local
coordinates $({q}^i)$, $1\leq i\leq \dim {Q}$ and that $\{\xi_a\}$ is a basis of ${\mathfrak g}$. The corresponding sections of the adjoint bundle  are the left-invariant vector fields $\xi_a^L$:
\[
\xi_a^L(g)=T_{\mathfrak e} L_g(\xi_a)
\]
where $L_g: G\longrightarrow G$ is  left translation by $g\in G$.
If
\[
A\left(\frac{\partial}{\partial {q}^i}_{( {{q}}, {\mathfrak e})}\right)=A_i^a \xi_a
\]
then  the corresponding horizontal lifts on the trivialization $U\times G$ are the vector fields
\[
\left(\frac{\partial}{\partial {q}^i}\right)^h=\frac{\partial}{\partial {q}^i}-A_i^a\xi_a^L\; .
\]

The elements of the set
\[
\left\{\left(\frac{\partial}{\partial {q}^i}\right)^h, \xi_a^L\right\}
\]
are by construction $G$-invariant, and therefore,  constitute  a local basis of sections $\{ e_i, e_a\}$ of ${\rm Sec} (T\bar{Q}/G)={\mathfrak X}(Q)\oplus {\rm Sec}( \widetilde{\mathfrak g})$.

Denote by $(q^i, y^i, y^a)$ the induced local coordinates of $T\bar{Q}/G$. Then
\[
B\left(\frac{\partial}{\partial {q}^i}_{( {{q}}, {\mathfrak e})},\frac{\partial}{\partial {q}^j}_{( {{q}}, {\mathfrak e})}\right)=
B_{ij}^a\xi_a
\]
where
\[
B_{ij}^c=\frac{\partial A^c_i}{\partial q^j}-\frac{\partial A^c_j}{\partial q^i}-{\mathcal C}^c_{ab}A^a_iA^b_j\; ,
\] the ${\mathcal C}_{ab}^c$ being the structure constants of the Lie algebra. The structure functions of the  Lie algebroid $T\bar{Q}/G\rightarrow Q$ are determined  (see \cite{{LMM-2005}}) by
\begin{eqnarray*}
\lcf e_i, e_j\rcf_{T\bar{Q}/G}&=& -B_{ij}^c e_c\\
\lcf e_i, e_a\rcf_{T\bar{Q}/G}&=&{\mathcal C}_{ab}^cA^b_i e_c\\
\lcf e_a, e_b\rcf_{T\bar{Q}/G}&=&{\mathcal C}_{ab}^c e_c\\
\rho_{T\bar{Q}/G}(e_i)&=&\frac{\partial}{\partial q^i}\\
 \rho_{T\bar{Q}/G}(e_a)&=&0\, ,
 \end{eqnarray*}
and for a Lagrangian function $L: \rk\times\stackrel{k}{\oplus}T\bar{Q}/G\longrightarrow \r$  the Euler-Lagrange field equations are
 \begin{eqnarray*}
 \frac{d}{dt^A}
 \left(\frac{\partial L}{\partial y^i_A}\right)&=&\frac{\partial L}{\partial q^i}
 +B_{ij}^c y^j_C\frac{\partial L}{\partial y^c_C}-{\mathcal C}_{ab}^cA^b_i y^a_C\frac{\partial L}{\partial y^c_C}\\
 \frac{d }{d t^A}\left(\frac{\partial L}{\partial y^a_A}\right)&=&
 {\mathcal C}_{ab}^cA^b_i
   y^i_C\frac{\partial L}{\partial y^c_C}
 -{\mathcal C}_{ab}^cy^b_C\frac{\partial L}{\partial y^c_C}\\
 0&=&\frac{\partial y_A^i}{\partial t^B}-\frac{\partial y_B^i}{\partial t^A}\\
 0&=&\frac{\partial y_A^c}{\partial t^B}-\frac{\partial y_B^c}{\partial t^A}-B^c_{ij}y^i_By^j_A
 +{\mathcal C}_{ab}^cA^b_i
 y^i_By^a_A+{\mathcal C}_{ab}^cy^b_Ay^a_B\; .
 \end{eqnarray*}

  If $Q$ is a single point, that is, $\bar{Q}=G$, then $T\bar{Q}/G={\mathfrak g}$,  the Lagrangian is a function
  $L: \rk\times\stackrel{k}{\oplus}{\mathfrak g}\longrightarrow \r$, and  the field equations  reduce to
 \begin{eqnarray*}
  \frac{d }{d t^A}\left(\frac{\partial L}{\partial y^a_A}\right)&=&
  -c_{ab}^cy^b_C\frac{\partial L}{\partial y^c_C}\\
  0&=&\frac{\partial y_A^c}{\partial t^B}-\frac{\partial y_B^c}{\partial t^A}+{\mathcal C}_{ab}^cy^b_Ay^a_B
 \end{eqnarray*}
   a local form of the Euler-Poincar\'e equations in field theory (see, for instance,  \cite{CGR-2001} and \cite{Mart-2004}).

\section*{Acknowledgments}

We acknowledge the partial financial support of {\sl Ministerio de
Innovaci\'{o}n y Ciencia}, Project MTM2007-62478, MTM2008-00689, MTM2008-03606-E/ and project
Ingenio Mathematica(i-MATH) No. CSD2006-00032
(Consolider-Ingenio2010) and S-0505/ESP/0158 of  the CAM.

\end{document}